\documentclass{aastex}

\usepackage[english]{babel}
\usepackage[T1]{fontenc}
\usepackage{graphicx}
\usepackage{amssymb,amsthm,amsmath}
\usepackage{ae}   
\usepackage{aecompl} 
\usepackage{rotating}
\usepackage{natbib}

\newcommand{\R}{{R}}

\newcommand{\overbar}[1]{\mkern 1.5mu\overline{\mkern-1.5mu#1\mkern-1.5mu}\mkern 1.5mu}

\renewcommand{\deg}{^\circ}

\newcommand{\pkdgrav}{{\tt pkdgrav}}
\newcommand{\Mercurysix}{{\tt Mercury6}}

\newcommand{\AU}{\,\mathrm{AU}}
\newcommand{\km}{\,\mathrm{km}}
\newcommand{\meter}{\,\mathrm{m}}
\newcommand{\cm}{\,\mathrm{cm}}

\newcommand{\yr}{\,\mathrm{yr}}

\newcommand{\degrees}{\,\mathrm{deg}}
\newcommand{\hours}{\,hours}
\newcommand{\hour}{\,\mathrm{h}}

\newcommand{\second}{\,\mathrm{s}}

\newcommand{\kg}{\,\mathrm{kg}}

\newcommand{\aei}{($a$,$e$,$i$ )}
\newcommand{\DSH}{D_{SH}}
\newcommand{\Dc}{D_{cluster}}
\newcommand{\Dp}{D_{pair}}
\newcommand{\ie}{{\it i.e.}}
\newcommand{\eg}{{\it e.g.}}
\newcommand{\etc}{{\it etc.}}
\newcommand{\wps}{\ensuremath{w_{\rm P1}}}
\newcommand{\PSone}{\protect \hbox {Pan-STARRS1}}
\newcommand{\oforder}{\mathcal{O}}

\def\aei{\mbox{($a$,$e$,$i$)}}

\begin{document}

\title{Properties and evolution of NEO families\\
created by tidal disruption at Earth}

\author{
Eva Schunov\'a\altaffilmark{1,2} (eva.schunova@gmail.com)
Robert Jedicke\altaffilmark{2},\\
Kevin J. Walsh\altaffilmark{3},
Mikael Granvik\altaffilmark{4},
Richard J. Wainscoat\altaffilmark{2}
Nader Haghighipour\altaffilmark{2}
}

\slugcomment{60 Pages, 16 Figures, 0 Table}

\altaffiltext{1}{Department of Astronomy, Physics of the Earth and
  Meteorology, Comenius University, Mlynsk\'a dolina, Bratislava, 942
  48, Slovakia}

\altaffiltext{2}{Institute for Astronomy, 2680 Woodlawn Dr, 
  Honolulu, HI, 96822, USA}

\altaffiltext{3}{Southwest Research Institute,1050 Walnut St, Suite 300, 
  Boulder, CO 80302, USA}

\altaffiltext{4}{University of Helsinki, Department of Physics,
  P.O. Box 64, 00014 Finland}

\shorttitle{The SFD of NEOs and ARM targets detected by \PSone\.}

\shortauthors{Schunov\'a \etal}

%
%
%
%
%


\begin{abstract}

We have calculated the coherence and detectable lifetimes of synthetic
near-Earth object (NEO) families created by catastrophic disruption of
a progenitor as it suffers a very close Earth approach.  The closest
or slowest approaches yield the most violent `s-class' disruption
events where the largest remaining fragment after disruption and
reaccumulation retains less than 50\% of the parent's mass.  The
resulting fragments have a `string of pearls' configuration after
their reaccummulation into gravitationally bound components
\citep{rich1998}.  We found that the average absolute magnitude ($H$)
difference between the parent body and the largest fragment is $\Delta
H \sim 1.0$.  The average slope of the absolute magnitude ($H$)
distribution, $N(H)\propto10^{(0.55\pm0.04)\,H}$, for the fragments in
the s-class families is steeper than the slope of the NEO population
\citep{mainzer2011} in the same size range.  The families remain {\it
  coherent} as statistically significant clusters of orbits within the
NEO population for an average of $\bar\tau_c = (14.7\pm0.6)\times10^3$
years after disruption.  The {\it detectable} lifetimes of tidally
disrupted families are extremely short compared to the multi-Myr and
-Gyr lifetimes of main belt families due to the chaotic dynamical
environment in NEO space --- they are {\it detectable} with the
techniques developed by \citet{fu2005} and \citet{Schunova2012} for an
average duration ($\bar\tau_{det}$) ranging from about 2,000 to about
12,000 years for progenitors in the absolute magnitude ($H_p$) range
from 20 to 13 corresponding to diameters in the range from about 0.5
to 10$\km$ respectively.  The maximum absolute magnitude of a
progenitor capable of producing an observable NEO family
(\ie\ detectable by our family finding technique) is $H_{p,max} = 20$
(about $350\meter$ diameter).  The short detectability lifetime
explains why zero NEO families have been discovered to-date.
Nonetheless, every tidal disruption event of a progenitor with
diameter greater than $0.5\km$ is capable of producing several million
fragments in the $1\meter$ to $10\meter$ diameter range that can
contribute to temporary local density enhancements of small NEOs in
Earth's vicinity.  We expect that there are about 1,200 objects in the
steady state NEO population in this size range due to tidal disruption
assuming that one $1\km$ diameter NEO tidally disrupts at Earth every
2,500~years.  These objects may be suitable targets for asteroid
retrieval missions due to their Earth-like orbits with corresponding
low $v_{\infty}$ which permits low-cost missions.  The fragments from
the tidal disruptions evolve into orbits that bring them into
collision with terrestrial planets or the Sun or they may be ejected
from the solar system on hyperbolic orbits due to deep planetary
encounters.  The end-state for the fragments from a tidal disruption
at Earth have $\sim5\times$ the collision probability with Earth
compared to the background NEO population.

\end{abstract}
\maketitle

\clearpage

\section{Introduction}
\label{sec:introduction}

In this work we evaluate some of the physical and observable
properties of tidally disrupted NEO families including their dynamical
evolution and coherence time scales, and their observable lifetimes
with modern asteroid surveys.  Our eventual aim is to set a limit on
the production rate of tidally disrupted NEO families for comparison
with theoretical expectations.

When an asteroid makes a close approach to within about the Roche
limit of a planet it may suffer morphological modifications due to
tidal forces depending on the circumstances of the encounter and the
asteroid's physical construction.  Inside the Roche limit\footnote{The
  Roche distance limit from a planet's center
  \citep{Roche1849,chandra1969} is given by $r_{Roche} = 1.52 (
  M_{pl}/\rho)^{1/3} = 2.46 R_{pl} (\rho_{pl}/\rho)^{1/3}$ where
  $M_{pl}$, $R_{pl}$ and $\rho_{pl}$ denote the mass, radius, and
  density, of the encountered planet respectively while $\rho$
  represents the density of the object making the encounter.} a
self-gravitating synchronously rotating liquid satellite orbiting a
spherical planet has no stable equilibrium
\citep{Roche1849,chandra1969} The structural changes may be as modest
as lifting layers of surface material \citep[\eg][]{nesvorny2010}, or
as violent as splitting the asteroid into a binary
\citep[\eg][]{walsh2008} or tearing it completely apart in the manner
of comet Shoemaker-Levy-9 \citep[\eg][]{asph1996}.  The tidal
disruption of a progenitor by a terrestrial planet is one of the most
likely creation scenarios for NEO families due to the relatively high
close encounter probability of the NEOs with those planets
\citep{fu2005}, however, there are as-yet no statistically robust
dynamical NEO families \citep{Schunova2012}.

Our claim that there are no known NEO families purposefully ignores
the well-known dynamical association between some meteor showers and
their cometary or asteroidal sources because they do not correspond to
the traditional usage of the `family' term that implies an orbital
relationship between much larger objects. Similarly, it has been
suggested that the Apollo asteroid (3200) Phaethon, the widely
recognized progenitor of the Geminid meteoroid stream
\citep[\eg][]{Whipple1983,jewitt2010}, is dynamically related to several other
NEOs such as 2005 UD \citep{ohtsuka2006,jewitt2006} and 1999 YC
\citep{ohtsuka2008}. This set of asteroids is thought to be the
remnant of a comet disintegration. The connection between Phaethon and
2005 UD in particular was proposed as a result of the similarity of
their orbits, their past dynamical evolution, and the fact that both
are optically blue which is uncommon for NEOs.  But the dynamical
association was not confirmed in the rigorous statistical
analysis of \citet{Schunova2012} --- the orbital similarity between the
two objects is not statistically significant.

Despite the lack of evidence for tidally disrupted NEO families in the
observed NEO population the existence of double craters on the
surfaces of Venus, Earth and Mars \citep{mel1991}, multiple crater
chains (known as `catenae') on the Moon and other natural satellites
\citep{mel1994}, and the bizarre shapes of some NEOs as revealed by
delay-Doppler radar images \citep[\eg][]{ostro1995,bot1999} imply that
close encounters and tidal disruption events are common in the NEO
population since binaries and multiple systems, a `string of pearls'
fragment distribution, and strange shapes, are all expected outcomes
of the tidal disruption process \citep[\eg][]{Rich2002}.

The different morphologies of the tidal disruptions are in part due to
the NEOs being relatively fragile `rubble-pile' asteroids composed of
the jumbled and re-accumulated remains of the aftermath of many
previous disruptions \citep[\eg][]{rich1998,pravec2000}. Further
evidence of the rubble-pile structure of asteroids is the sharp limit
in their spin period distribution at $P\sim2.2$\hours\ for
asteroids with $H<22$ (larger than about $200\meter$ diameter).  The
period corresponding to the rotation barrier agrees well with the
theoretical rotation rate limit for objects with a density of
$\sim$2.7~g\,cm$^{-3}$ \citep{har1996,pravec2000}.  Asteroids that
spin faster than the observed limit experience a centrifugal
acceleration on the surface that exceeds the local gravitational
acceleration leading to mass loss through spalling or complete
rotational disruption of the object.  

{\it In situ} evidence for the
asteroids's rubble pile constitution come from spacecraft missions
to the NEOs (243) Mathilde and (25143) Itokawa by the NEAR
\citep[\eg][]{yeomans1997} and Hayabusa spacecraft
\citep[\eg][]{fujiwara2006} respectively which revealed that their inner structure
must be porous.  Direct imaging from the
spacecraft revealed obvious rubble pile structures on their surfaces,
with loosely distributed boulders and gravel, and crater morphologies
that can only be explained if they are due to impacts on a weak or
fragmented target \citep[\eg][]{green1994,love1996}.  Perhaps
conclusively, \citet{asph1996} showed that the tidal disruption of
comet Shoemaker-Levy-9 (SL9) by Jupiter matches the observed
disruption event if the progenitor had a rubble pile construction.

Mild tidal events might be also capable of producing meteoroid streams.
The close passage of an object by a planet can lift material from the
surface that then spreads along the object's orbit with time due to
the small $\Delta v$ imparted by the lifting process and gravitational
perturbations by other objects in the solar system
\citep{kornos2009,toth2011}.  After hundreds of years the particles
will be distributed along the entire orbit and will produce meteors
and fireballs if they enter Earth's atmosphere. However, due to the
single-event nature of the tidal disruption the stream would not
be replenished and the putative meteoroid stream could only be observed
for a short time after creation. \citet{toth2011} calculated the
maximum activity of a meteor shower originating from the
putative Pribram meteorite progenitor to be only 1 meteor in 8 days
visible from a single observing place.

\citet{rich1998} performed numerical simulations of NEO distortion and
disruption by Earth's tides using more realistic asteroid shapes, spin
rates, axis orientations, perigee distance $q$, relative speed at
infinity $v_{\infty}$ and body orientation at periapse, and identified
four classes of tidal encounter outcomes (that should not be confused
with the capital letters used to identify asteroid taxonomic classes):
\begin{description}
\item[s)] {\bf catastrophic disruptions} in which the largest
  remaining fragment retains less than 50\% of the parent's mass
  (similar to SL9),
\item[b)] {\bf rotational breakup} in which the largest remaining
  fragment retains 50\%-90\% of the parent's mass,
\item[m)] {\bf mild disruption} in which less than 10\% of the
  parent's original mass is lost and,
\item[n)] {\bf no mass loss} but possible morphological modification.
\end{description}

They found that the result of the disruption is primarily determined
by the relative speed at infinity $v_\infty$ and periapse distance
$q$. The most violent s-class disruptions occur at small periapse
distances at low speeds because this configuration allows the objects
to spend more time in Earth's proximity and consequently offers more
time for the tidal forces to act on the body.  In a simplified case of
an NEO rapidly rotating in the prograde direction (\ie\ when the angle
between the object's rotational pole and the north ecliptic pole is
$<90\arcdeg$), centrifugal acceleration and tidal forces act together
to cause mass shedding and/or disruption so that the range of $q$ and
$v_{\infty}$ where the disruptions occur widens as the rotation rate
increases.  On the other hand, retrograde rotation reduces the
severity of disruption and mass loss and may even prevent the
disruption \citep{rich1998}.

While the state of tidal disruption modeling is mature and continues
to improve both in theory and in the lab, comparatively little has
been done on estimating the rate at which disruptions actually take
place and understanding the evolution of the fragments after the
disruption process.  An NEO with $D>1\km$ ($H_p\lesssim17.5$)
suffers an s-class event near Earth 
or Venus roughly every 3,200
years and some kind of disruption event (M-, B- and s-class) occurs
once about every 1,000 years \citep{rich1998}.  More recently \citet{toth2011}
calculated that the frequency of tidal disruptions of any kind
for NEOs with $D>200\meter$ ($H_p\lesssim21$) is
$\sim1.6\times10^{-4}$/year or roughly one tidal disruption every
6,200 years --- within a factor of 2 of the \citet{rich1998} disruption
rate.  Our eventual goal is to measure or set a limit on the
disruption rate in an effort to distinguish between the two calculations and thereby test the entire NEO orbit and disruption modeling process.

If the disruption rate is high enough, and the fragments remain
dynamically distinguishable as a tidally disrupted family long enough,
then it should be possible to identify them in the NEO population.
However, recent attempts to identify families in the NEO population
all yielded zero candidates
\citep[\eg][]{drummond2000,fu2005,Schunova2012}.  While several NEO
clusters were identified they could not be shown to be statistically
significant.  The main goals of this work are to determine the
coherence and detectable lifetimes of tidally disrupted NEO families
and use this information to set an observational limit on the number
of tidally disrupted NEO families.

The tidal disruption of an NEO could pose an increased impact threat to
Earth above and beyond that calculated from the current NEO models
\citep[\eg][]{bot1994,bot2002a,ivanov2008} that do not account for
local enhancements in the orbit element phase space due to families.
Since asteroids tend to repeat their trajectories and encounters, if
an asteroid is tidally disrupted at Earth then it is likely that the
fragments will also return to Earth's environs. A limit on the
frequency of disruption events that lead to the creation of NEO
families will thus provide the necessary information to assess their
effect on the NEO collision rate with Earth.

\section{Method}
\label{sec:methods}

We carried out $N$-body simulations of the tidal disruptions of
roughly spherical gravitationally-bound `rubble-pile' asteroids on a
range of Earth-crossing orbits and their subsequent gravitational
reaccumulation into secondary objects.  The secondaries form a tidally
disrupted NEO family that we integrate forward in time for $10^7\yr$
to inspect their orbital evolution, and applied a cluster
identification algorithm \citep{Schunova2012} at logarithmically
time-spaced intervals to measure the families' coherence time
(\ie\ the time during which they can be detected as a cluster of
objects on similar orbits). We then employed the \PSone\ Moving Object
Processing System \citep{denneau2013} to simulate observational
selection effects and calculated the NEO families' detectable
lifetimes. Finally, we measured the families' size-frequency
distribution and the porosity of the family members.

\subsection{Simulating tidal disruptions at Earth}
\label{sec:TIDI-simulations}

We used the method developed by \citet{gra2012} and implemented in the
OpenOrb software package \citep{gra2009b} to generate 10,000 NEOs on
Earth encountering orbits with flat distributions in all six orbit
elements with semi-major axes in the range $0.5\AU<a<7.4\AU$,
eccentricities with $0<e<1$, inclinations ($i$) from $0\deg$ to
$90\deg$, and angular elements spanning $[0\arcdeg,360\arcdeg)$.  In
  this way our results can be easily normalized to match any desired
  NEO model. Initial epochs for the tidal-disruption simulations were
  randomly chosen within one synodic lunar period (29.530589 days)
  starting on MJD 55461 (2010 Sep 22). The randomization over all
  possible geometries between the earth, Moon and Sun would require
  extending the initial epoch distribution over the 19-year Metonic
  Cycle but we assumed that our choice will have a negligible effect
  on our results because the dynamics of NEOs are dominated by close
  planetary encounters on time scales that are much longer than the
  Metonic cycle or synodic lunar period.  We then down-selected the
  NEOs to the 718 objects with $v_{\infty} \le 35\km\second^{-1}$ and
  $q \le 2.5\,R_{\earth}$, the limiting values enveloping tidal
  disruption events near Earth \citep{rich1998}, and performed N-body
  simulations of their passage through the Earth-Moon system.

The synthetic NEOs are rubble pile models
consisting of about 2,000 solely gravitationally-bound identical rigid
$120\meter$ diameter `unit' spheres.  Thus, the smallest possible
fragment in our simulations is a single unit sphere with absolute
magnitude $H_0=22.2$ assuming a geometric albedo\footnote{Roughly the
  median of measured albedos of all asteroids from {\tt
    http://sbn.psi.edu/pds/resource/albedo.html} and widely used in
  the astronomical community.} of $\rho = 0.15$.  The unit spheres
were arranged in Hexagonal Closest Packing
\citep[HCP,][]{Leinhardt2000} to form roughly spherical progenitors
with diameters of $\sim 2\km$.  However, to a large extent, the tidal
disruption process is independent of the dimensions of
the unit spheres and the progenitor 
\citep{solem1996} and we are free to scale the unit
spheres as needed to simulate both larger and smaller objects.

The simulations assigned every unit sphere a density of
$\rho=3,340\kg\meter^{-3}$ and yielded progenitors with a bulk density of
$\sim2,130\kg\meter^{-3}$ and porosity $\Phi_p \sim 37$\% at the realized
average packing efficiency of $\sim63$\%.  These values are in good
agreement with measurements of rubble-pile NEOs with similar
dimensions such as (25143)~Itokawa \citep{abe2006}, the binary
(66391)~1999~KW$_4$ \citep{ostro2006}, and the triple NEO 1994~CC
\citep{brozovic2011}.

While the size of a synthetic progenitor is not particularly important
in the disruption simulation its shape and rotation period are
critical.  Oblate and prolate objects are more easily disrupted than
spherical ones and faster rotation periods also enhance the tidal
disruption process.  The complete parameter space determining tidal
disruption outcome was examined in earlier works
\citep[\eg][]{rich1998, walsh2006} but we want to set an upper limit
on the tidal disruption rate while at the same time not wasting CPU
time.  We balanced these objectives by using roughly spherical
progenitors generated with the HCP technique but using a median spin
rate of $P=4.3\hour$ \citep{pravec2002} for NEOs with H$<$22 and a
disruption-favorable prograde spin-axis.  The overall effect should be to enhance
the tidal disruption rate of our progenitors above the actual rate
such that we can set an upper limit.

Every simulation was set in the
planetocentric frame with 50,000 time steps of 5 seconds each
corresponding to roughly 3\,days for modeling the entire disruption
process.  The starting distance was 20 Earth radii from Earth, so the
fragments typically traveled roughly 40 Roche radii ($\sim$ 100 Earth
radii) during the simulation time and by the end of the simulation
they were far away enough from Earth that tidal forces were
negligible.  From that time in the simulation we numerically
integrated the particles under the forces of all the planets in the
solar system (see \S\ref{sec:Num-integrations} below).

Fig.~\ref{fig:Vinf-q-success-trials} illustrates that the 120
disrupted objects all have $v_{\infty} \la 35.0\km\second^{-1}$ and $q
\la 2.5\,R_{\earth}$ and it is evident that the disruptions
become more violent as each parameter decreases (\ie\ the progenitor
suffers more loss of material).  The most violent s-class disruptions
take place exclusively when the object approaches Earth within
$2\,R_\earth$, less than half the distance to geostationary orbit, and
at speeds $\le 15\km\second^{-1}$, consistent with the results of
\citet{rich1998} and \citet{walsh2006} (as expected since we used
their software for our simulations).

A total of 120 of the 718 tidal disruption simulations produced more
than two fragments --- the rest were either gentle events with no
disruption or binary-producing events.  We ignored the binaries
because it has already been established that identifying genetically
linked NEO pairs in a statistically significant manner is exceedingly
difficult \citep{Schunova2012}.  As expected, the majority (73) of the
disruptions were of the mild m-class, while roughly half as many
progenitors (32) suffered the deep encounters required for s-class
disruption.  In the remainder of this paper we usually
consider only the 32 s-class disruptions.

\begin{figure}
 \centering
 \includegraphics[width=0.7\textwidth]{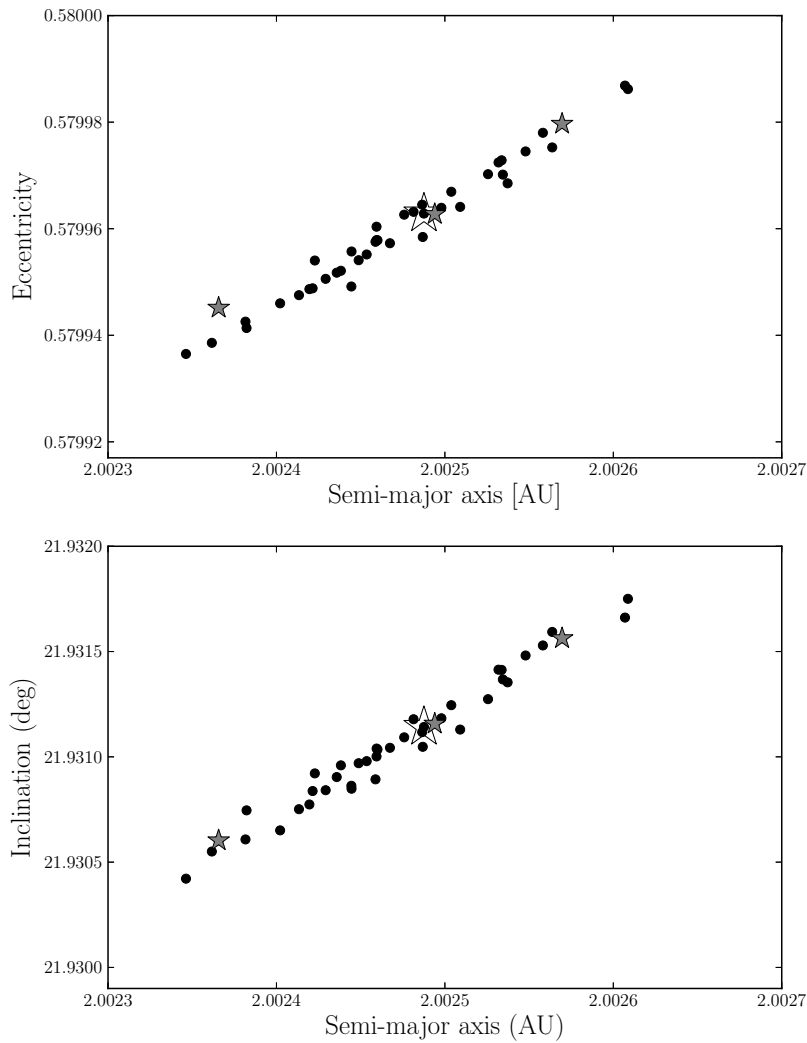}
 \caption{Post-disruption configuration of members of a synthetic NEO
   family created in a s-class tidal disruption. (top) eccentricity vs. semi-major axis and (bottom) inclination vs. semi-major axis.  The large white star
   represents the position of the largest fragment and the filled gray
   stars represent the positions of the next three largest members of
   the family.  `Post-disruption'
means the end of the \pkdgrav\ tidal disruption simulation when we begin our
n-body integration of all the fragments under the influence of all
major bodies in the solar system.  The simulations typically end when the fragments
are beyond $\sim100\R_\earth$.}
\label{fig:Post-disruption-fragment-configuration}
\end{figure}

The $\sim27$\% of our simulations that generated s-class tidal
disruptions created what we call s-class families in which the
progenitor is shattered and the fragments are configured in
a `string of pearls' tidal stream (see
Fig.~\ref{fig:Post-disruption-fragment-configuration}).    The members of the family have very similar orbital elements and
a power-law size-frequency distribution (SFD) where the mass is distributed
along the string rather then concentrated at its center.  We think
that s-class families will eventually be detected in the NEO
population and this work determines their dynamical and detectable
lifetime.
 
\begin{figure}
  \centering
  \includegraphics[width=0.7\textwidth]{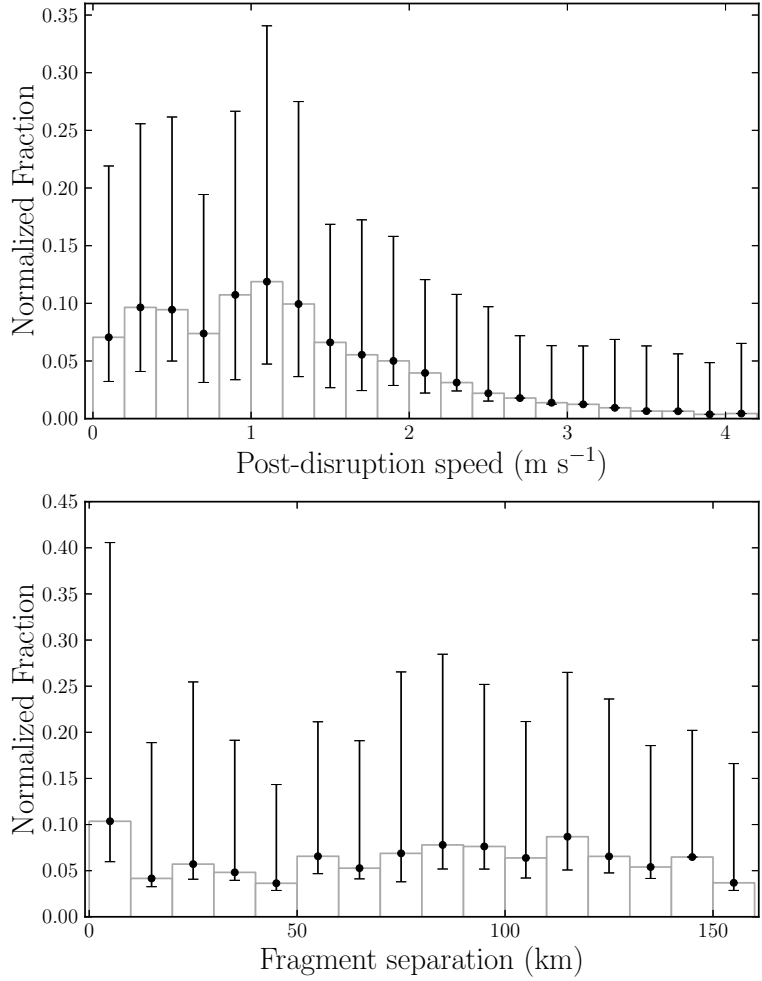}
  \caption{Post-disruption speed and separation of members of
    synthetic s-class NEO families created by tidal disruption near
    Earth when the fragments are well beyond Earth's Roche limit at the
end of the n-body disruption simulation (typically beyond $\sim100\R_\earth$).  Both values are always with respect to the largest remnant
    in each family.  The error bars represent the asymmetric RMS
    variation of the fractional distribution in the bin across all the
    synthetic s-class families.}
\label{fig:V-R-ej-distribution}
\end{figure}

The maximum post-disruption speed ($v_{pd}$) of $\sim
4\meter\second^{-1}$ (see Fig.~\ref{fig:V-R-ej-distribution}) is much
less than speeds typical of impact-generated catastrophic disruptions
in the main belt that can reach up to half the impact speed of the
projectile (the typical impact speed in the main belt is $\sim
5\km\second^{-1}$, \citet[\eg][]{bot2001, scheeres2002}).  The average
post-disruption speed of s-class family members is
$v_{pd}\sim1.2\meter\second^{-1}$ --- less than the escape speed for a
$2\km$ rubble pile asteroid.  The average speed is about $10\times$
higher than predicted by \citet{toth2011} who calculated that the
speed of boulders escaping the surface of a rubble pile NEO during
close Earth encounters is $\gg\sim 1\meter\second^{-1}$.

It is interesting that the fragments can escape even though their
relative velocities with respect to the largest fragment at tidal
disruption were less than the escape speed (see
Fig.~\ref{fig:V-R-ej-distribution}).  They can do so because tidal
disruptions are gentle events during which tidal effects continue to
gently accelerate the fragments away from each other well after the
primary disruption event.  In the case of NEO tidal disruptions at
Earth the initial disruption generates an average of about
$0.4\meter\second^{-1}$ relative speed between the fragments that
increases asymptotically almost another 50\% to about
$0.6\meter\second^{-1}$ at the end of the simulation.  By the end of
the simulation the fragments are no longer gravitationally bound
because there are several tens to hundreds of kilometers separation
between them and it is appropriate to continue the simulations with a
long-term dynamical orbit integrator (see
\S\ref{sec:Num-integrations}).

\subsection{NEO tidal disruptions at other terrestrial planets}
\label{sec:Mars-Venus}

\begin{figure}
\centering
\includegraphics[width=1\textwidth]{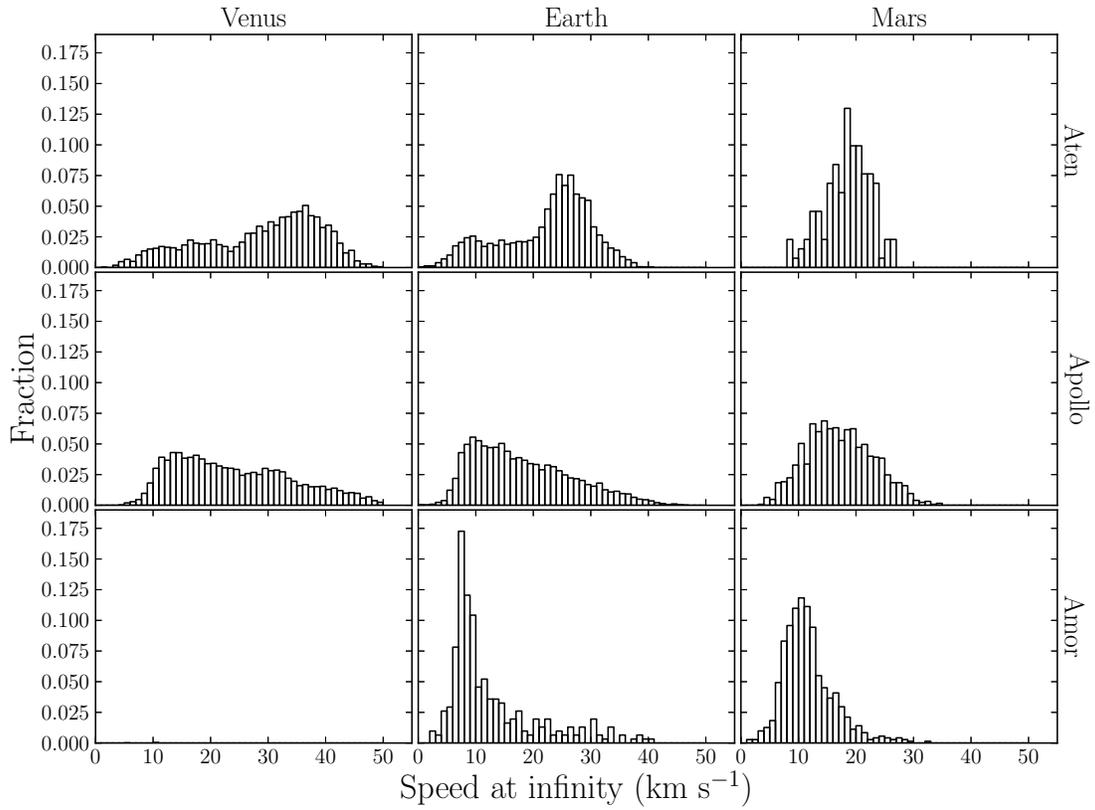}
\caption{Distributions of the relative $v_\infty$ for Amor,
  Apollo and Aten groups at Venus, Earth and Mars. Amor asteroids
  never approach Venus (lower left panel) but their velocity
  distributions with Mars and Earth peak at small values favorable to
  tidal disruption.}
\label{fig:vel.distrib.neas}
\end{figure}

We excluded all the other terrestrial planets and the Moon from
consideration because they are not expected to contribute a
significant flux of tidally disrupted families in NEO phase space.
This is mostly due to the distribution of $v_{\infty}$ for the NEO
dynamical sub-groups at the planets as illustrated in
Fig.~\ref{fig:vel.distrib.neas}. 

Mercury is relatively easy to exclude as a source of s-class NEO
families because of 1) its small mass 2) the low spatial density of NEOs that
can approach it and 3) their high encounter speeds.

Venus and Earth have almost $10\times$ larger collision probabilities
with NEOs than Mercury and Mars \citep{bot1994}.  We estimate that
Venus is about 1/3 as effective as Earth at producing s-class
disruption families because 1) about 42\% of NEOs are on
Venus-crossing trajectories compared to $\sim 84$\% on Earth-crossing
orbits \citep{bot2002a} and 2), since NEOs are typically moving faster
near Venus than when approaching Earth, only about 56\% of them have
$v_{\infty}<30\km\,\second^{-1}$ with respect to Venus while $\sim
86$\% of the Earth-crossing NEOs have slow enough approach speeds to
allow effective tidal disuption (see Fig.~\ref{fig:vel.distrib.neas}).
Thus, Venus contributes to the tidally disrupted NEO family population
at about 33\% the rate contributed by Earth.

It is possible that Mercury and Venus could disrupt `rogue' asteroids
that are not NEOs ($\eg$ objects on orbits that are entirely interior
to Earth's orbit \citep[\eg\ ][]{zavodny2008} that could make a close passage
by either Mercury or Venus and the fragments could then evolve into
NEO phase space) but this possibility is small and, in any event, we
will show below that the detectable lifetime of tidally disrupted NEO
families is short.  Thus, it is unlikely that tidally disrupted
families created in this manner will stay coherent and evolve into NEO
space in a time frame that will allow them to be detectable.

The Moon was excluded because its smaller Roche sphere makes it much
less efficient than Earth at disrupting rubble piles \citep{rich1998}
(the ratio of the cross-sectional area of the Moon's Roche sphere to
Earth's is only $\sim0.074$).

Finally, despite Mars being only 1/10$^{th}$ Earth's mass it 1) is
much closer to the Main Belt, 2) orbits the Sun in a region with a
much higher flux of asteroids and 3) NEO encounter speeds are slower
than at Earth which would yield an increased tidal disruption
probability --- so it is not immediately obvious whether it will
create more or less s-class NEO families than Earth.  The mean impact
rate of $1\km$ diameter NEOs on Mars is predicted to be about $10\times$
smaller than the impact rate of the same-size NEOs on Earth
\citep{bot1994} and, since the impact rate on Earth is about one
$1\km$ diameter NEO every $\sim6\times10^5$ years
\citep[\eg][]{stuart2003}, the rate on Mars is about once every $\sim
6 \times 10^6$ years. The interval between close encounters with Mars
sufficient for s-class disruption roughly scales 
as the cross-sectional area of its Roche sphere compared to the
planet, so that the s-class disuption rate is still on the Myr time
scale.

To estimate the number of NEO families created near Mars by tidal
disruption it is also necessary to account for the Intermediate source
Mars-crossers (IMC) that feed the NEO population \citep{bot2002a}.
The IMCs are to Mars as NEOs are to Earth --- they are the subset of
the Mars-crossing asteroid population that borders the main belt with
orbital parameters $q>1.3\AU$, $2.06\AU\le a \le 2.48\AU$ or
$2.52\AU\le a < 2.8\AU$, with $i\lesssim 15\deg$, and a combination of
\aei\ values such that they cross the orbit of Mars during a secular
oscillation cycle of their eccentricity \citep{bot2002a,mig1998}.
\citet{michel2000} estimates that there are about $4 \times 10^4$ IMCs
with $D>1$ km with an average lifetime of $\tau_{MC}\sim29$\,Myr and,
during their lifetime, about 4\% of IMCs suffer encounters within 3
Mars radii of Mars\footnote{W.~F.~Bottke, personal communication} ---
about 1 close approach every $\sim18,000$ years. Our Mars tidal
disruption simulations suggest that there is $\sim$17\% probability of
a close Mars encounter resulting in a s-class tidal
disruption so we estimate that there is one family created by tidal
disruption every 100,000 years.  Since this time is considerably
longer than the predicted disruption rates at Earth
\citep[\eg][]{rich1998,toth2011}, and because the most likely s-class
disruptions at Mars are of IMCs, not NEOs, we exclude Mars-induced
tidal disruptions in the NEO population from consideration.

\subsection{Post-disruption dynamical integrations}
\label{sec:Num-integrations}

The geocentric position and velocity of the tidal disruption fragments
at the last time step (MJD=55499) were converted back to their heliocentric values
and numerically integrated forward in time for 60~Myr.  All the
integrations used the N-body Burlisch-Stoer algorithm from the
\Mercurysix\ software package \citep{cham1999} with variable time
steps that allows good tracking of the orbital evolution during close
encounters with planets. The initial timestep was set to 3 days. 

\begin{figure}
\centering
\includegraphics[width=\textwidth]{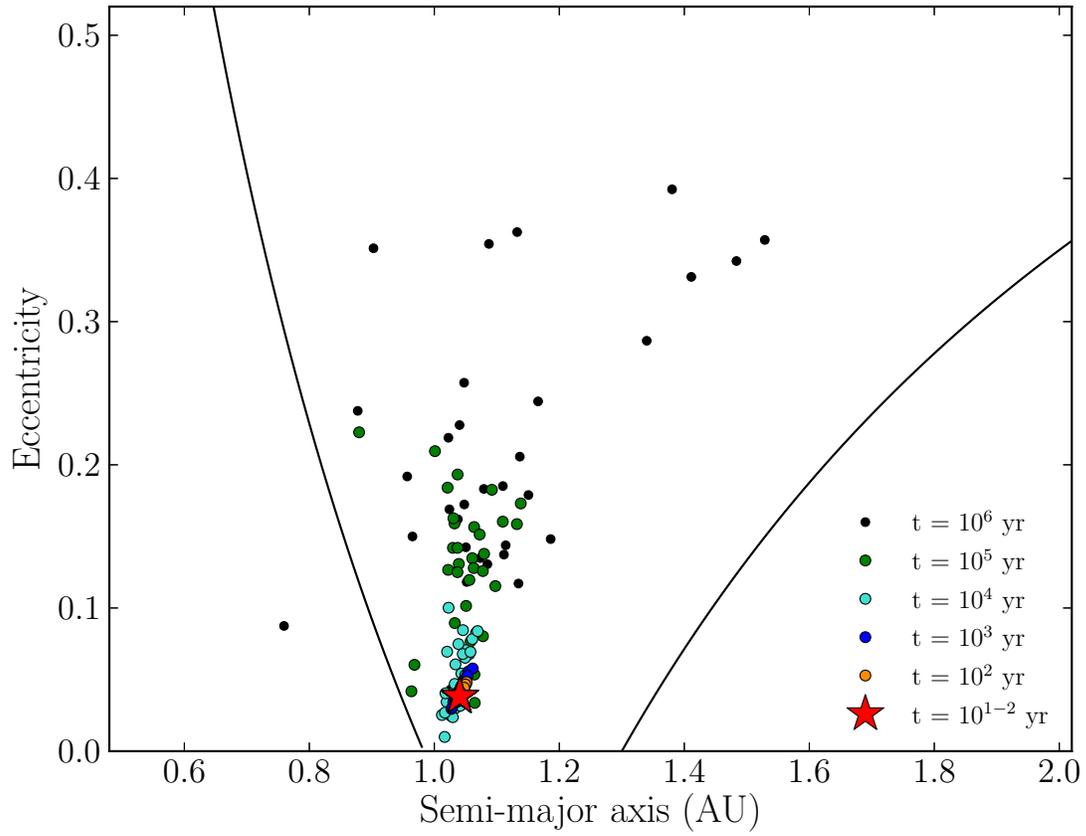} \\
\caption{Time evolution of eccentricity and semi-major axis of the
  members of a single s-class NEO tidal disruption.  This family is
  detectable for 70,000 years using the method described in
  \S\ref{sec:IdentifyingNEOFamilies}.}
\label{fig:EM-evolution-example-ae}
\end{figure}

The orbital time evolution of the members of a s-class family as
illustrated in Fig.~\ref{fig:EM-evolution-example-ae} proceeds rapidly
as expected for an object in the inner solar system on a
planet-crossing orbit. The progenitor has an orbit with perihelion
near Earth and the orbits of the fragments remain practically
unchanged during the first few hundred years in which individual
fragments become evenly distributed along the original orbit
(\ie\ throughout the whole range of mean anomaly) forming a possible
meteoroid stream \citep[\eg][]{toth2011}.  In this particular
family the members then evolve to higher eccentricity over the next
10s of thousands of year and only then begin to spread in semi-major
axis.  All the fragments remained as NEOs throughout the first
$10^6\yr$.

We included only gravitational forces in our integrations because NEOs
of the size of our smallest fragments with $H=23$ ($D\sim100\meter$)
are influenced by non-gravitational forces on timescales much longer
than the coherence time scale of the tidally disrupted NEO families.
The most recently measured average Yarkovsky drift rate for km-scale NEOs
with albedos measured by WISE is $\sim 10^{-9}\AU$/year
\citep{nugent2012} with the rate being a few times higher for
$100\meter$-scale objects.  Our measured average `dynamical drift
rate' for the tidally disrupted fragments in our simulations was about
$1,000\times$ higher \ie\ the average rate of evolution of the
fragments just under the gravitational influence of the major objects
in the solar system.  Since the non-gravitational effects are minimal
even on the smallest objects in our simulation we are justified in
ignoring their effect on all the objects.

\subsection{Simulating observational selection effects}
\label{sec:MOPS}

We need to account for selection effects in the detection of the
fragments to assess the ability to detect NEO families created through
catastrophic tidal disruption.  Observational selection effects will
favor brighter objects at any $\aei$ values and surveys will
preferentially discover those objects on low-inclination orbits for
which the objects approach Earth closely and slowly
\citep[\eg][]{Jedicke2003}.  The latter effect favors the detection of
tidal disruption fragments because objects on those types of orbits
are precisely those that are most likely to be disrupted.  The former
effect makes tidally disrupted NEO families more difficult to detect
because a small but detectable progenitor will disrupt into smaller
fragments for which not enough members are detectable by modern
surveys.

We simulated observational selection effects representative of the
entire known NEO population using the Pan-STARRS Moving Object
Processing System \citep[MOPS,][]{denneau2013} and the full sample of
250,000 NEOs from the Synthetic Solar System Model (S3M)
\citep{gra2011}.  We used the 47 lunation survey configuration of
\citep{gra2009a} because they showed that it is a good proxy for all
surveys that have discovered NEOs to-date and roughly reproduces the
number of known NEOs and their orbit distribution.  The fact that the
simulation does so implies that it can roughly simulate the
observational selection effects imprinted on the known NEO population
and on the synthetic members of tidally disrupted NEO families in our
study.

The PS1 survey and the NEO simulation are described in detail in the
last three references but, briefly, the \PSone\ system
\citep{kaiser2010} has a $7\degrees^2$ field of view and is able to
reach a limiting magnitude of $V_{lim} \sim 21.7$ with their
\wps\ filter. To mimic the known NEO population the NEO simulation
used a limiting magnitude of $V_{lim} \sim 22.7$ (following
\citet{gra2009a}) and included both opposition fields and small solar
elongation fields \citep[morning and evening `sweet
  spots';\ ][]{Chesley2004}) covering a region of up to
$5,500\degrees^2$ per lunation. Each field is visited twice on three
different nights within a lunation with a minimum of four days between
the visits. To mimic weather, 25\% of the nights are randomly
excluded. When an object is bright enough to be detected MOPS links
detections on the same night into `tracklets'.  This is followed by
inter-night linking of the tracklets into `tracks' that are then
tested for consistency with detections of an object on a heliocentric
orbit.  Objects with tracklets on at least three nights spread over
the course of 7-10 days will yield a good orbit determination with
\PSone's astrometric accuracy of about $0.1\arcsec$
\citep{milani2012}.

We ran the orbits for all 1720 fragments from the 32 s-class tidal
disruptions at 10 different epochs (0, 1, 10, 100, 10k, 20k, 50k, and
100k years after disruption) through the 47 lunation MOPS simulation
to determine their detectability.  We used a fixed $H=0$ for all the
fragments simply to determine if and when the objects appeared in the
simulated survey and their apparent $V$ magnitudes at those times.
\ie\ fragment $i$ has apparent magnitude $V_i(t,0)$ at time $t$ with
$H=0$.  We could then assign any absolute magnitude $H_i$ to a
fragment and determine its apparent magnitude in the MOPS simulation
as $V_i(t,H_i)=V_i(t,0)+H_i$ \citep[\eg][]{harris1998}.  This allowed
us the flexibility of assigning any SFD to the families including
varying the size of the progenitor, the size of the largest fragment,
and the slope of the distribution.  Once the apparent magnitude was
calculated for each fragment in each field we then determined whether
it was above the simulated \PSone\ limiting magnitude ($V_{lim} \sim
22.7$) to determine if the fragment was detected.  There had to be at
least two detections of the same (moving) object per night to form a
tracklet and multiple tracklets of the same object over several nights
were required for the orbit determination.  We did not account for
trailing losses in the detection algorithm because the losses are
relatively small for the typical rates of motion of the fragments in
the survey ($<1\degrees$/day).

The quality of orbit determination is crucial for assessing the
statistical significance of NEO clusters \citep{Schunova2012}.  Thus,
in this work we applied stricter criteria for NEO detection than the
actual \PSone\ survey for which the first NEO detections are
complemented by follow-up observations at other observatories.  We
required that the arc length (in time) for an NEO family member be
$\ge90$ days and that the object be detected with $\ge3$ tracklets in
at least one lunation with additional tracklets in other lunations.
With these requirements in place the orbital uncertainties are reduced
to the level that makes it viable to search for NEO clusters in the
known population comparable to the family search of
\citet{Schunova2012}.

\subsection{Identifying NEO families}
\label{sec:IdentifyingNEOFamilies}

We adopted the cluster identification method developed by
\citet{fu2005} that we used previously to search for families in the
known NEO population \citep{Schunova2012}.  The method relies on
identifying tight clustering of family members' orbit elements using
the \citet{sou1963} orbital similarity criterion ($\DSH$) that
incorporates all 5 Keplerian orbital elements.

The cluster identification method is described in detail by
\citet{Schunova2012} and \citet{fu2005} but we provide a short primer
here.  The technique recognizes that when families form and evolve in
orbit element space the entire family may fit within the `envelope' of
the cluster but individual members may form a set of tightly linked
pairs spanning a sub-set of the envelope.  As time passes the envelope
increases in size, the pairs become more widely spaced and the links
between them are `broken' as the density of the members in orbit
element space decreases.  The envelope of fragments is identified as a
cluster in the 5-dimensional orbit element space because they all have
mutual $\DSH<\Dc$ (the $\DSH$ cluster threshold).  The tightly linked
sub-set is identified as the set of member pairs with mutual
$\DSH<\Dp$ where $\Dp<\Dc$.  The largest set of paired fragments that
all have $\DSH<\Dp$ is called the `string' (because they are all
connected with $\DSH<\Dp$).  Finally, every candidate NEO cluster must
fulfill two conditions:
\begin{itemize}
\item String size to cluster size ratio $SCR>SCR_{min}$
\item Pair fraction $PF>PF_{min}$
\end{itemize}
where the string and cluster sizes are the number of objects in the
string and cluster respectively and the pair fraction is the actual
number of identified pairs divided by the total number of possible
pairs in the cluster.

In our earlier work we showed that $\Dc=0.06$, $\Dp=0.05$,
$SCR_{min}=0.75$ and $PF_{min}=0.5$ allowed statistically significant
identification of clusters of objects with $>4$ members
\citep{Schunova2012}. One 4 member cluster was identified in the NEO
population but only at the 2- to 3-$\sigma$ significance level.

\section{Results}
\label{sec:Results}

\subsection{Orbit distribution of precursor bodies to tidally-disrupted NEO families}
\label{sec:HeliocentricOrbitalEvolution}

\begin{figure}
  \centering
  \includegraphics[width=1.0\textwidth]{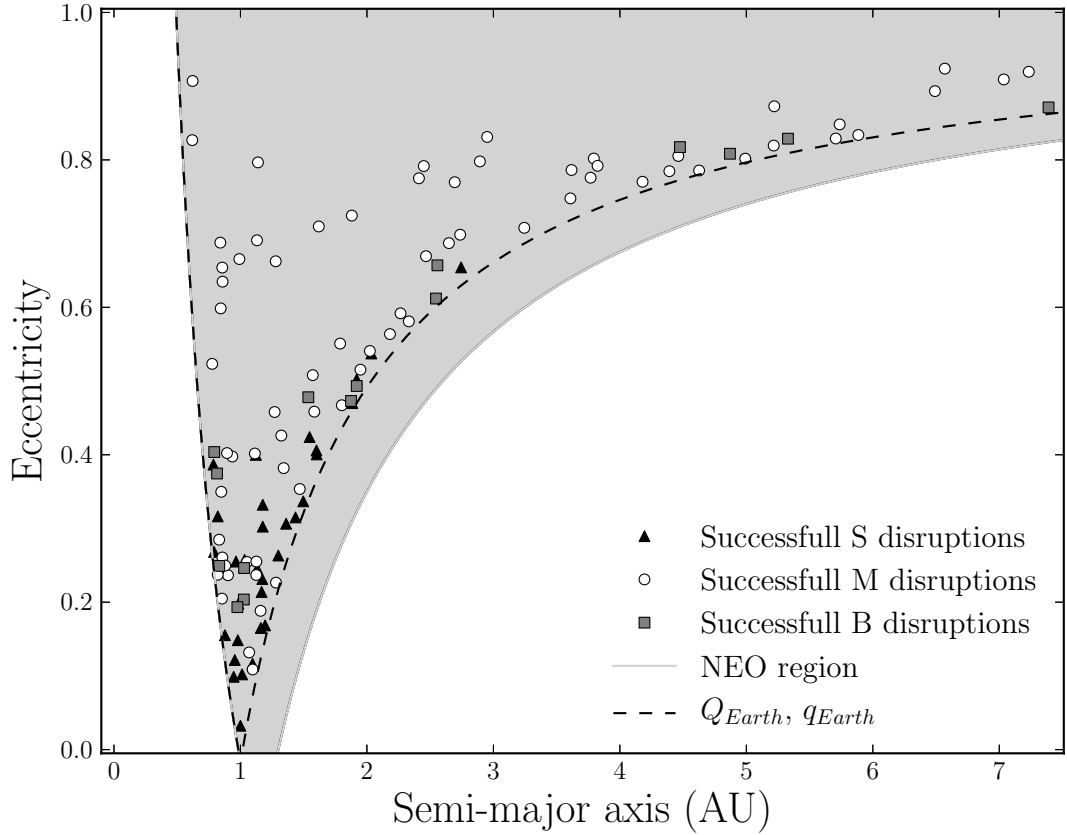}
  \caption{Eccentricity vs. semi-major axis of NEO progenitors that
    underwent M, B and s-class disruptions. The NEO region is depicted
    by the shaded area.  The left and right dashed lines that converge
    at $a\sim1$\,AU correspond to orbits having aphelion and
    perihelion at Earth's perihelion and aphelion distance
    respectively.}
\label{fig:TIDI-avse-success-trials}
\end{figure}

\begin{figure}
  \centering
  \includegraphics[width=0.9\textwidth]{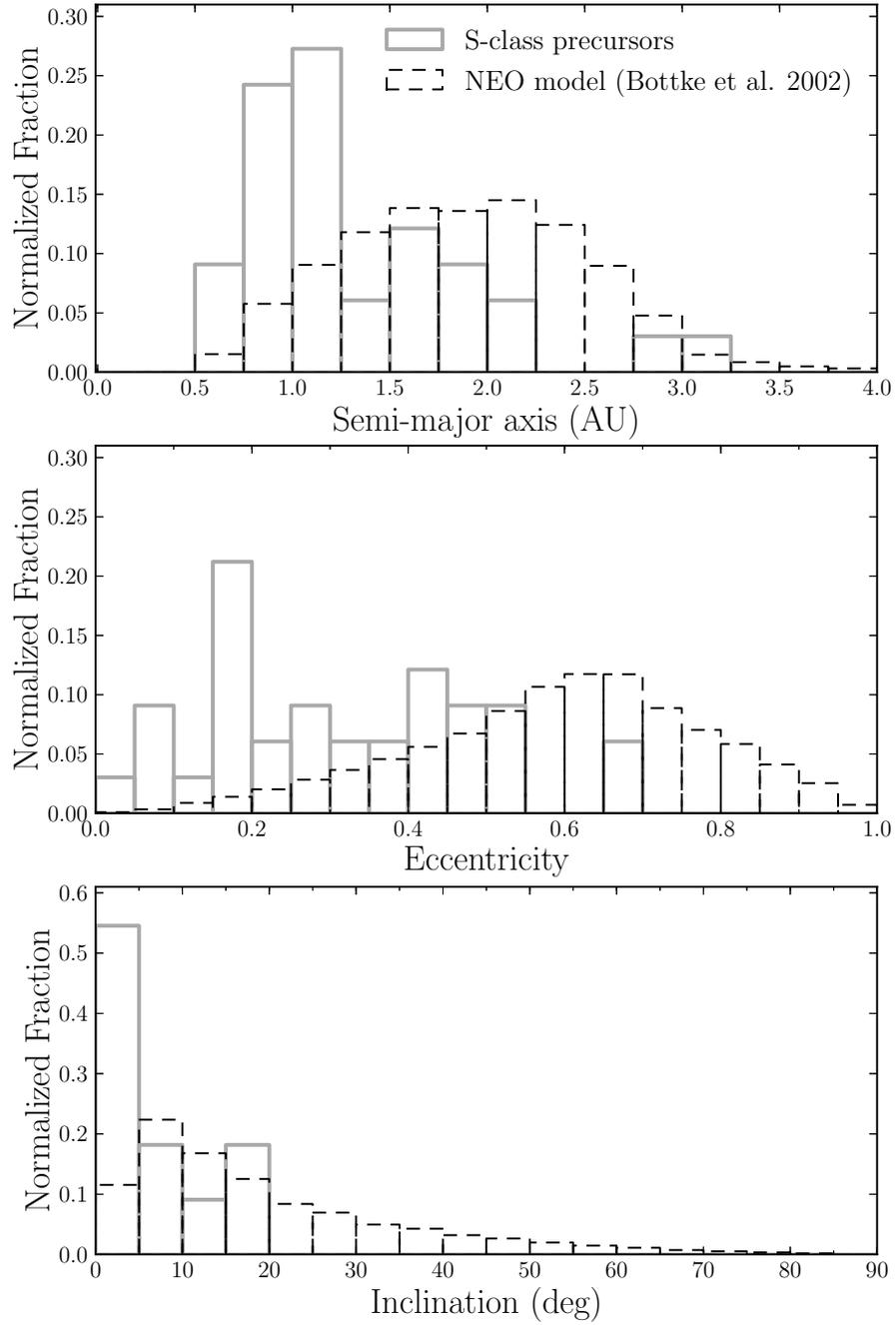}
  \caption{Fractional distribution of semi-major axis, eccentricity
    and inclination for NEOs from the \citet{bot2002a} NEO model
    (dashed) and NEO progenitors (solid grey) that result in s-class tidal
    disruptions by Earth.}
\label{fig:Parent-NEO-aei-distrib}
\end{figure}

Pre-encounter orbits that result in tidal disruptions usually have
aphelion or perihelion distances that lie near Earth's orbit and the
s-class disruptions of interest in this work are more tightly
clustered along the Earth aphelion and perihelion lines than the M-
and b-class disruptions (Fig.~\ref{fig:TIDI-avse-success-trials}).  More generally, the s-class
disruption probability is enhanced for objects on orbits with
$0.5\AU<a<1.25\AU$, $e<0.5$ and $i<5\deg$ (Fig.~\ref{fig:Parent-NEO-aei-distrib}) so that the progenitors' orbit distribution is a subset of the entire NEO population \citep{bot2002a}.  This is a result of the
fact that these types of orbits are more tangential to Earth's orbit
and the objects therefore spend more time in Earth's environs where
they may be subject to disruption.  Note that more than half of all
the s-class disruptions occur in the $i<5\deg$ bin.  This is
interesting because it is now generally thought that the
\citet{bot2002a} NEO model under-represents the fraction of the
population on these low-inclination orbits by about a factor of 2 in the sub-km size range
\citep{mainzer2011}. Thus, there may be more NEOs on orbits subject to
s-class disruptions than this figure suggests.

A possible explanation for the enhancement of low-inclination NEOs
\citep[\eg][]{mainzer2011,Rabinowitz1993,greenstreet2012} is that it is
{\it because} of a recent tidal disruption event that has increased
the population of objects on Earth-like orbits in the NEO population.
The enhanced number of objects in Earth-like orbits will increase the
 Earth-impact rate above current estimates and would
increase the likelihood of the Earth capturing temporary satellites
\citep{gra2012}.

\subsection{Porosity \& size-frequency distribution\\of tidally disrupted NEOs}
\label{sec:TIDI-SFD}

The size-frequency distribution (SFD) of the fragments in a tidally
disrupted NEO family is critical to the determination of its
detectability.  \citet{Schunova2012}'s NEO family detection method
requires that $>4$ known members be brighter than a limiting
absolute magnitude $H_{max}=23$.  In our simple model the number of objects in
the family brighter than the limit depends on 3 factors: the absolute
magnitude of the progenitor ($H_{p}$), the difference between the
absolute magnitudes of the largest fragment and the progenitor
($\Delta H_1$), and the slope $\alpha$ of the SFD.  The cumulative SFD
for the fragments is then
\begin{equation}
N(H) = 10^{\alpha (H-H_1)} = 10^{\alpha ( H - H_p - \Delta H_1 )}
\end{equation}
\ie\ there are $N(H')$ fragments with $H<H'$ in the family with the
largest fragment having an absolute magnitude of $H_1 = H_p + \Delta
H_1$.  We used the results of our s-class tidal disruption simulations
to determine the values of $\Delta H_1$ and $\alpha$.

\begin{figure}
\centering
\includegraphics[width=\textwidth]{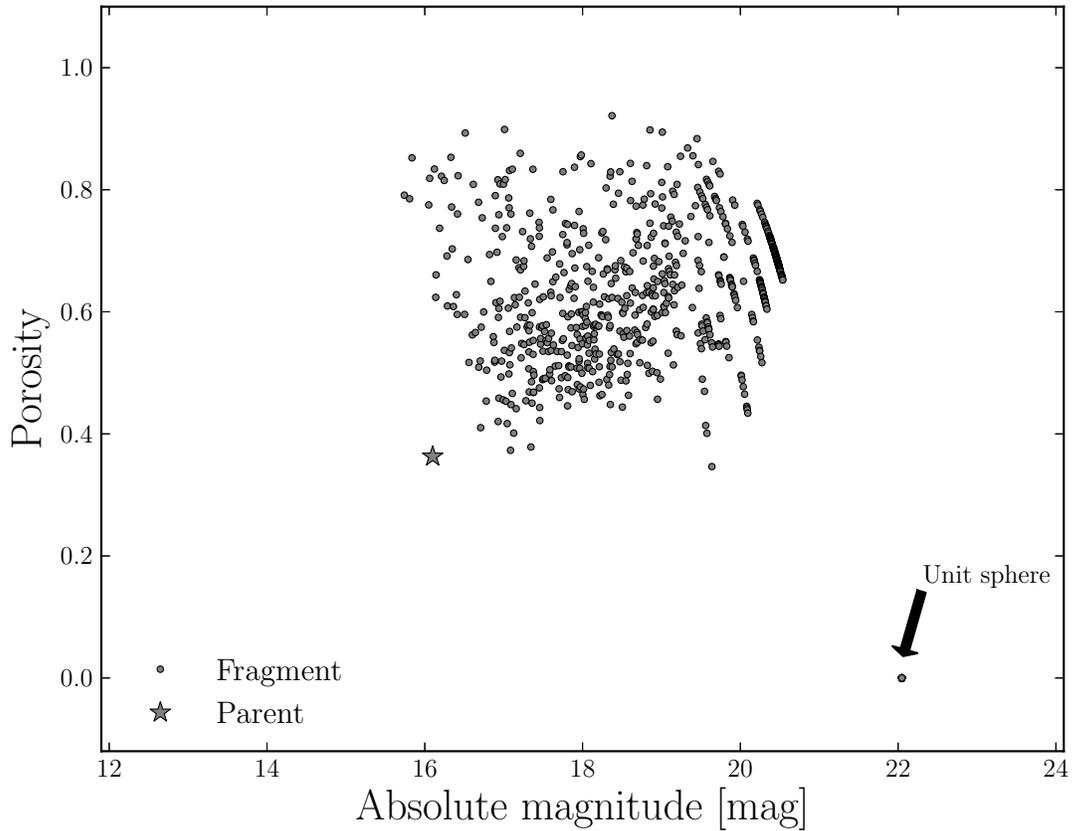} 
\caption{Porosity of s-class disruption fragments as a function of the
  absolute magnitude of the fragment. Some fragments have absolute
  magnitudes brighter than the progenitor, and most fragments have a
  higher porosity, due to a `fluffy' re-accumulation/re-arrangement of
  the unit spheres after disruption.  The points that lie along
  discretized curves near $H_{frag}\sim20$ are due to fragments built
  from a small number of unit spheres.  Fragments consisting of only
  one unit sphere have $\Phi=0$ as shown in the lower right.}
\label{fig:Sclass.porosity.vs.H}
\end{figure}

To calculate a fragment's absolute magnitude we assumed that it is
spherical, has an albedo of 0.15, and radius as provided by
\pkdgrav. This will overestimate the absolute magnitude because
\pkdgrav\ provides the dimension of the fragment's longest axis as the
`radius'.  For example, a `fragment' consisting of two unit spheres has a
`radius' equal to twice the unit sphere radius.  The problem is
illustrated in Fig.~\ref{fig:Sclass.porosity.vs.H} which shows that
the porosity ($\Phi$) of the fragments is always higher than the
porosity of the progenitor (by porosity we mean the macro-porosity ---
the fraction of the volume of an object occupied by voids).  The
discretized $\Phi$ values near $H_{frag}\sim20$ are due to fragments
that are built from a small number of unit spheres.  

The tidal disruption process first increases the porosity of the
progenitor as its equipotential surface is stretched towards Earth
while approaching perigee before disruption into an s-class family.
The disruption fragments continue to evolve and re-accumulate while
still under the influence of tidal tension thereby `fluffing' the
internal structure of the larger fragments by increasing the size of
internal void spaces.  The resulting high-porosity elongated fragments
might resemble the NEO (1620) Geographos
\citep[\eg][]{bottke1999,walsh2006} but these elongated and loosely
bound rubble pile fragments probably cannot survive long in this state.  Moderate-scale collisions or another series of close
encounters together with self-gravity will decrease the macro-porosity
by gradually collapsing the void spaces.

\begin{figure}
\centering
\includegraphics[width=\textwidth]{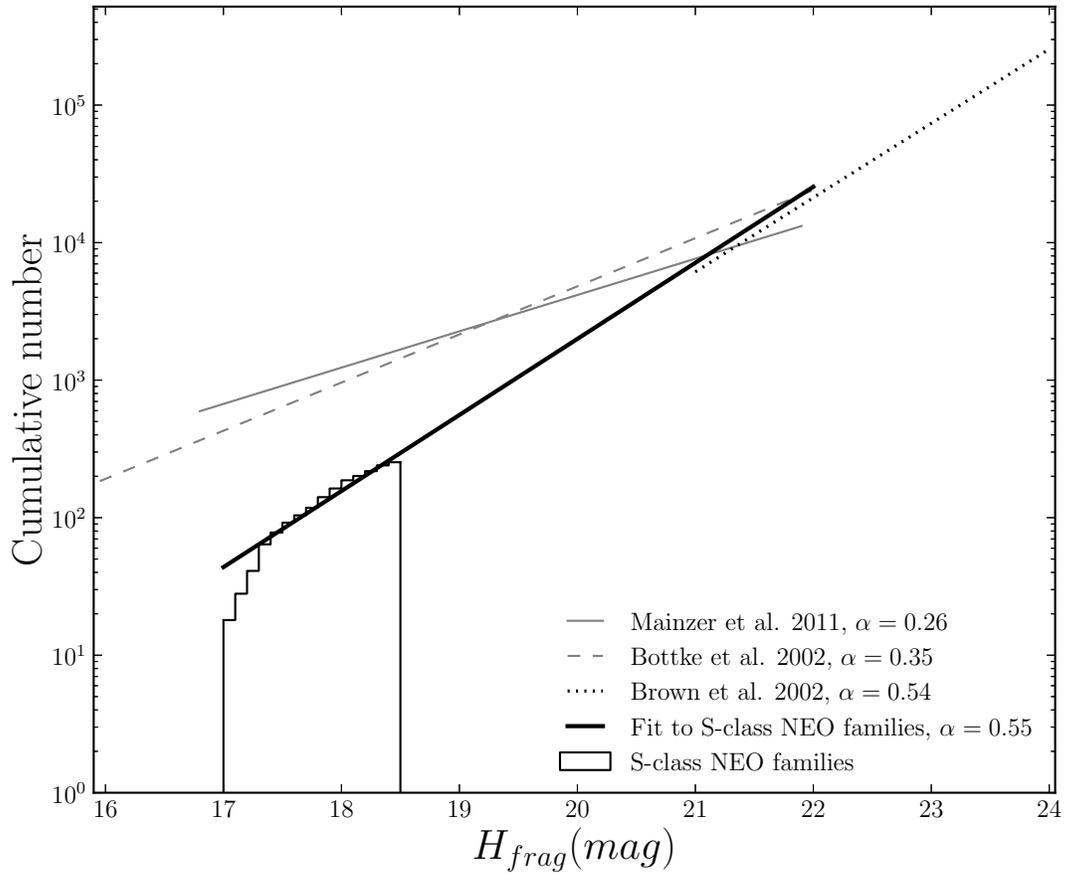} 
\caption{Three measured cumulative NEO SFDs for the real
  population \citep{mainzer2011,bot2002a,brown2002} and the SFD of our
  synthetic s-class disruption fragments with a fit as described in
  \S\ref{sec:TIDI-SFD}.}
\label{fig:SclassH.all.cumulative}
\end{figure}

The determination of the s-class family SFD slope ($\alpha$) from the
tidal disruption simulations was also problematic due to the
discretization of the unit spheres and the relatively small number of
fragments in each disruption.  To overcome the latter problem we
combined all our s-class disruption fragments.  Even then, we
could only fit the resulting SFD over a narrow range of absolute
magnitudes from the largest fragment ($H_1$) to $H^*$, the absolute
magnitude where the porosity (and mass and size) of the fragments
becomes discretized, because of a clear turnover in the SFD caused by
the unit sphere discretization.  We empirically determined $H^*$ by
starting with just the two largest family members and iteratively
adding more members as we monitored the slope of the SFD.  $H^*$ was
fixed at the value where the SFD began to decrease, $H^*=18.5$, at
which the fragment diameter is $\sim6\times$ that of a unit
sphere. This technique results in a cumulative SFD with a slope of $\alpha
= 0.55 \pm 0.04 \mathrm{(stat.)}$ (see Fig.~\ref{fig:SclassH.all.cumulative}).

Given that the calculated absolute magnitude of the fragments is
problematic, we estimated the systematic uncertainty in the slope by
recalculating each fragment's absolute magnitude in a `worst case'
scenario assuming that the fragments have the same porosity as the
progenitor instead of using \pkdgrav's diameter.  The average change
in absolute magnitude for the fragments was about $+0.7$\,mags
\ie\ the fragments became smaller and their absolute magnitudes
increased.  This technique yielded a slope of $\alpha = 0.66$ so we
took the systematic uncertainty on the slope to be $\oforder(0.11)$.

Our measured slope of $\alpha = 0.55 \pm 0.04 \mathrm{(stat.)} \pm
0.11 \mathrm{(sys.)}$ is steeper than the SFDs of the overall NEO
population of $\alpha = 0.35\pm0.02$ 
\citep[for $H<22.0$;\ ][]{bot2002a} and from WISE observations where
$\alpha=0.26\pm0.03$ \citep[for $H>17$;\ ][]{mainzer2011}.  On the other hand, our results are
consistent with the slope of $0.54\pm0.02$ measured from fireballs
\citep{brown2002} in the size range from $1\cm$ to $200\meter$
($21\lesssim H \lesssim 43$) and with the SFD of $\alpha \sim 0.56$
for surface boulders on the NEO (25143)~Itokawa in the 20\,cm to
$20\meter$ size range \citep[corresponding to $26\lesssim H \lesssim 36$;\ ][]
{saito2006}.  Furthermore, the measured slopes are in 
agreement with the theoretical value of $0.5$ for self-similar
collisional systems \citep[\eg][]{dohnanyi1969,obrien2005}.  The steep
slope for the s-class families is also consistent with the suggestion
of \citet{toth2011} that the tidal disruption of rubble-pile asteroids
may increase the SFD slope of the NEO population.

We do not consider the absolute magnitude (or porosity) `problem' to
be serious for four reasons.  First, while some of the fragments have
outrageously high porosities of $>90$\% the bulk of them lie in the
40-60\% range compatible with measured values for NEOs and main belt
asteroids. For example, (25143)~Itokawa has a porosity of about 40\%
\citep{fujiwara2006} while \citet{baer2011} showed that porosities of
main belt asteroids smaller than 300~km in diameter range from 40\% to
60\% or more with the highest measured asteroid porosity being 75\%.
Second, and as discussed in the last paragraph, our measured SFD slope
is consistent with the slope of small NEOs, expectations for fragments
from tidal disruptions, and the theoretical value for self-similar
collision cascades.  Third, in our analysis below
(\S\ref{sec:Lifetime}) we will use 0.5~mag bins in the progenitors's
absolute magnitude --- on the scale of the worst case error in the
calculated absolute magnitudes.  Thus, we will use a value of
$\alpha=0.55\pm0.11$ recognizing that the uncertainty is dominated by
the systematic errors in our analysis.  Finally, we note that the
spin-frequency distribution \citep{Warner2009} of all asteroids shows
a well-known discontinuity at about $100\meter$ that suggests a
transition from a rubble pile internal structure to monoliths at about
that diameter --- thus, it is possible that there really is an actual
discretization of asteroid interiors with a fundamental building block
size of $\sim100\meter$.

\subsection{Dynamical lifetime and coherence time of tidally disrupted s-class NEO families}
\label{sec:Coherence}

Several dynamical mechanisms operating in the NEO region are capable
of significantly increasing the orbital eccentricities and
inclinations of NEOs that eventually cause them to impact the Sun, a
planet, or be ejected from the solar system \citep{Gladman2000}.  NEOs can also evolve onto hyperbolic
orbits or suffer collisions with terrestrial planets.  We will compare
the end states of NEOs created in s-class tidal disruptions at the
Earth to the overall NEO population in \S\ref{sec:end-states} but we
measured their average dynamical lifetime to be
$\sim10$~Myr in agreement with \citet{Gladman2000} \ie\ the time at which
half of the tidally disrupted NEOs remain in the NEO population.  The
dynamical lifetime of the families sets an upper limit to the
families' coherence time.

\begin{figure}
\centering
\includegraphics[width=\textwidth]{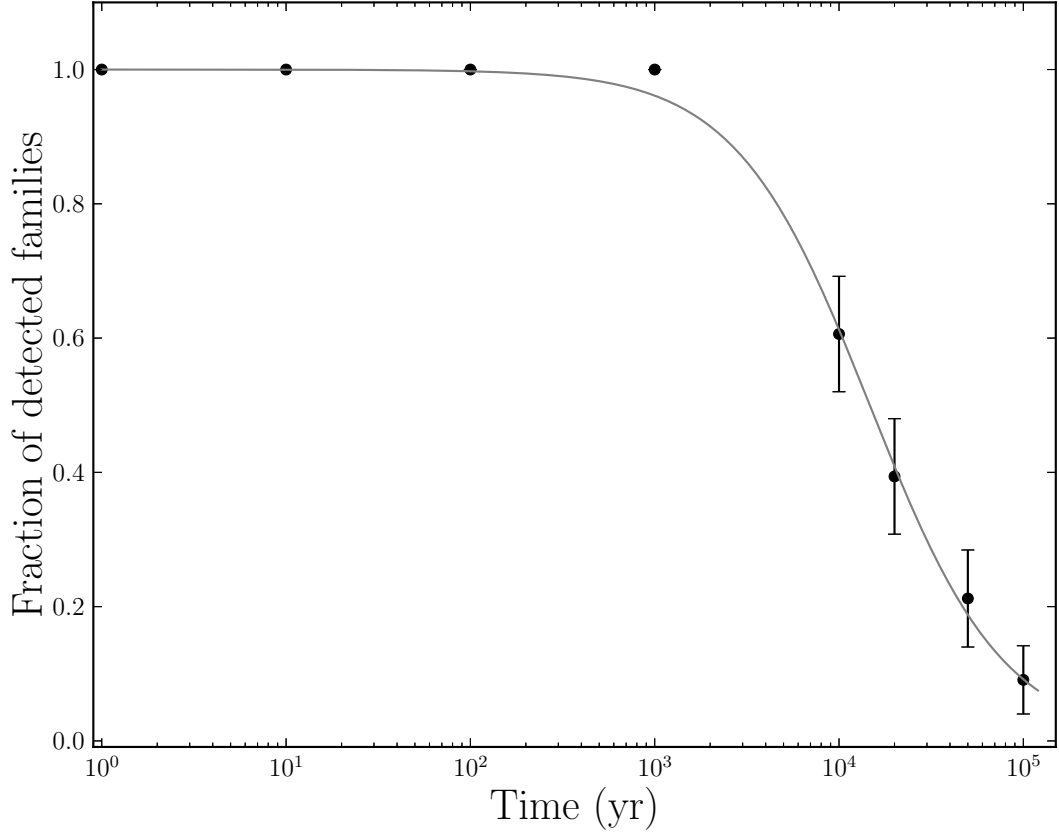} 
\caption{The fraction $f$ of the 32 s-class NEO families created by
  tidal disruption during a close encounter of their progenitors with
  Earth that are detectable as families as a function of time.  The
  data was fit to a function of the form
  $f(t)=\{1+\exp[(\log_{10}(t)-\log_{10}(\tau_c))/w_c]\}^{-1}$
  yielding a coherence time of $\tau_{c}=(14.7\pm0.6)\times10^3\yr$
  when 50\% of the families can be detected.}
\label{fig:Earth-fam-lifetime}
\end{figure}

We define the coherence time for an NEO family to be the time duration
during which the family members can be identified using our cluster
identification algorithm described in
\S\ref{sec:IdentifyingNEOFamilies} {\it without} taking into account
the fragments' detectability by astronomical surveys \ie\ when the
cluster identification is based solely on the similarity of the
members' orbital elements.  To measure the coherence time we extracted
the orbits of each member of each family from the numerical
integration at 10 epochs: at $t = 0\yr$, $t = 10^k\yr$ with $k=0-6$,
and at 20 and 50 kyr.  Then we applied our cluster identification
algorithm to each family at each epoch.  All the s-class
families can be identified as clusters for only the first 1,000~years
(Fig.~\ref{fig:Earth-fam-lifetime}) and the average coherence time of the
32 families is
$(14.7\pm0.6)\times10^3\yr$ (\ie\ the average time after disruption
for which $\ge 50$\% of the families remain dynamically coherent).
This time scale is extremely rapid compared to most of the known main
belt families with ages of $>10^6\yr$ or even the youngest family
associated with (1270)~Datura that has an age of
$\sim(450\pm50)\times10^3\yr$ \citep{nesvorny2006}.

The coherence time depends on the parameters of the cluster finding
algorithm which, in turn, were selected on their ability to
discriminate NEO clusters within the background NEO orbit element
population \citep{Schunova2012}.  An alternative cluster finding
algorithm may thus yield longer (or shorter) coherence times.  For
instance, the Earth's atmosphere acts as a `cluster finding algorithm'
in the sense that it can be used to identify meteoroid streams that
intersect Earth's orbit in the form of meteor showers.  The typical
meteoroid stream coherence time of roughly 1,000~years \citep[\eg][]{toth2011} is
$\sim10\times$ shorter than our measured value for tidally disrupted
NEO familes because the concept of the NEO family changes from
meteoric dust size particles to asteroid-sized particles, and because meteor
streams are strongly influenced by non-gravitational forces like the
Yarkovsky effect and radiation pressure.

\subsection{Detectable lifetime of tidally disrupted NEOs}
\label{sec:Lifetime}

The coherence time sets an upper limit on the
clusters' detectable lifetime --- the time during which the family
members may be discovered by current surveys --- because the coherence
time does not account for observational selection effects as described
in \S\ref{sec:MOPS}. Observational limitations will reduce the time
during which a family can be identified because some of its members
will not be detected due to their unfavorable observing geometry, rate
of motion, or faintness.  If there are not enough ($>4$) family
members above the limiting absolute magnitude the family itself is
simply undetectable using the \citet{Schunova2012} technique.  Thus,
we define the `detectability lifetime' ($\tau_{det}$) as the time
during which members of tidally disrupted NEO families can be
efficiently discovered by ground-based asteroid surveys and identified
as a statistically significant cluster in orbit element space.

\begin{figure}
\centering
\includegraphics[width=\textwidth]{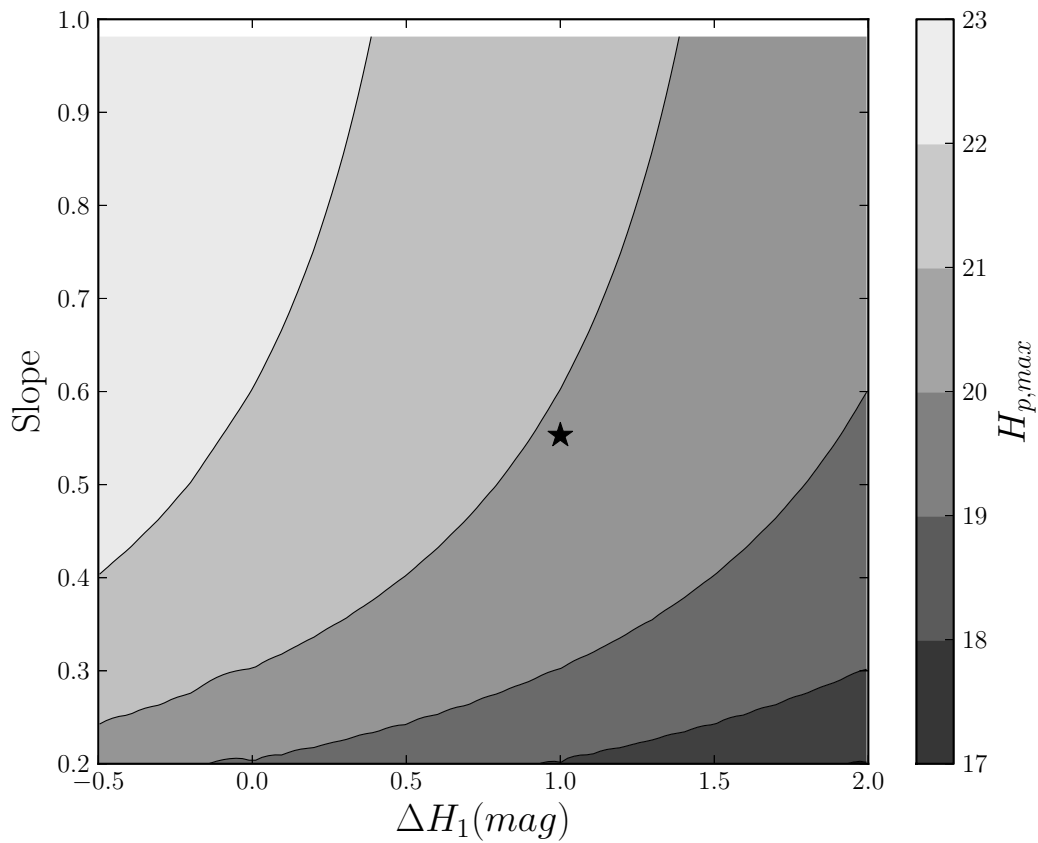} 
\caption{Maximum absolute magnitude ($H_{p,max}$) of a tidal
  disruption progenitor able to produce a detectable NEO family as a
  function of the fragments' size-frequency distribution slope
  ($\alpha$) and the difference between the absolute magnitudes of the
  progenitor and the largest fragment ($\Delta H_1$).  The star near the center represents our measured values from the \pkdgrav\ simulations of s-class disruptions (see \S\ref{sec:TIDI-SFD}).}
\label{fig:Hp-vs-slope-deltaH1}
\end{figure}

The smallest detectable progenitor's diameter (see
Fig.~\ref{fig:Hp-vs-slope-deltaH1}) must increase 1) as the difference
in size between the progenitor and largest fragment increases and 2)
as the SFD slope decreases.  Both results are as expected because
smaller fragments are more difficult to detect, and a shallow SFD
produces only a few large fragments above the limiting absolute
magnitude (\ie\ there might not be enough of them to be detected as a
statistically significant cluster with our method).  While
Fig.~\ref{fig:Hp-vs-slope-deltaH1} spans the full range of slopes and
$\Delta H_1$ from our tidal disruption simulations
(\S\ref{sec:TIDI-SFD}), the smallest detectable progenitor's diameter
(maximum absolute magnitude, $H_{p,max}$) of an `average' tidally
disrupted s-class family has $H_{p,max}\sim20$ (for
$\bar\alpha\sim0.55$ and $\overbar{\Delta H_1}\sim1.0$).  This
absolute magnitude corresponds to minimum progenitor diameters of
$\sim0.3\km$ and $\sim0.77\km$ assuming albedos of 0.2 and 0.03
typical of S- and C-class asteroids respectively \citep{mainzer2011}.

\begin{figure}
\centering
\includegraphics[width=0.9\textwidth]{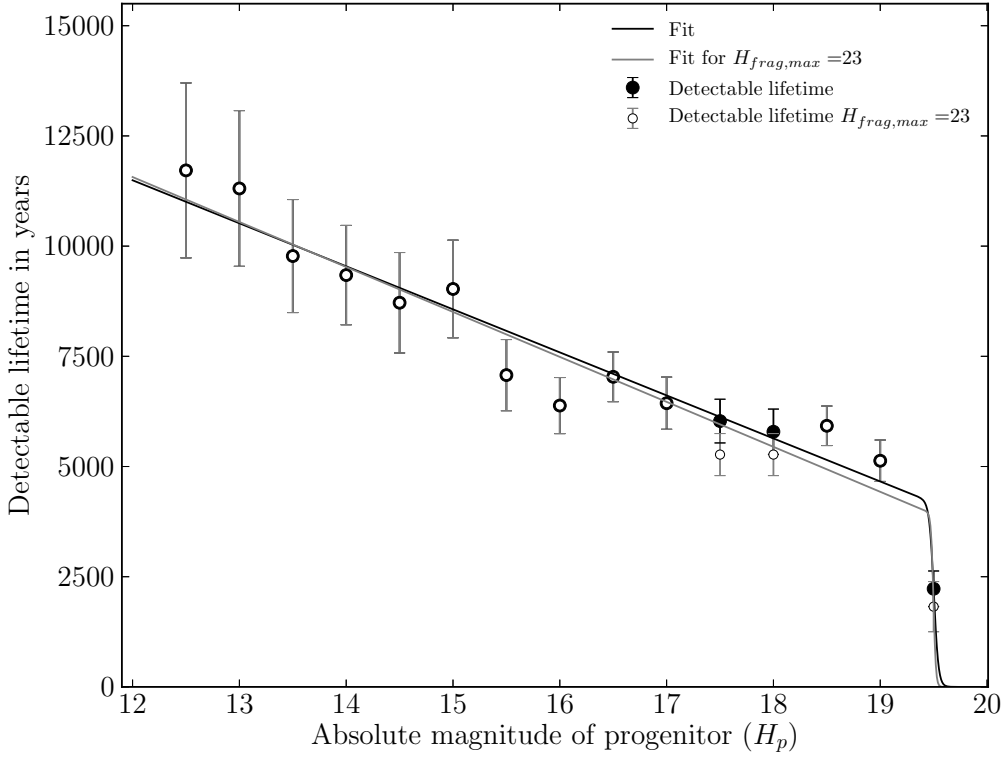} 
\caption{Detectable lifetime of tidally disrupted NEO families as a
  function of the progenitor's absolute magnitude $H_p$ with and
  without a cut in fragment size at $H_{frag,max}=23$.  The
  detectability lifetime includes observational selection effects
  typical of all the NEOs in the known population as well as the
  efficiency of the cluster detection algorithm.  The lifetime without
  the $H_{frag,max}$ cut was fit to a function of the form $\tau_{det}
  = (\tau_0 + m \, H_p ) / \{1 + \exp[{ (H_p - H_{limit})/ H_{width}}]\}$
  with $(\tau_0=23.8 \pm 1.5)\times10^3$~years, $m=(-1.0 \pm
  0.1)\times10^3$~years/mag, $H_{limit}=19.5$ and
  $H_{width}=0.1$. The fit with the $H_{frag,max}$ cut is
  almost identical. (We do not provide the
  uncertainties for the last two parameters because they are much
  larger than the precision required to reproduce the fit shown on the
  figure.)}
\label{fig:Obs-Lifetime-vs-Hp}
\end{figure}

We applied our cluster detection algorithm on the family members
detected in the survey simulation (\S\ref{sec:MOPS}) as we did to
calculate the coherence time (\S\ref{sec:Coherence}).  The
fragments were assigned absolute magnitudes according to the average
SFD with $\Delta H_1=$1 mag and $\alpha=0.55$.  We then measured the
detectable lifetime of all the families as a function of the
progenitors' absolute magnitude in 0.5~magnitude $H$ steps (Fig.~\ref{fig:Obs-Lifetime-vs-Hp}) for two cases:  1) all detectable
objects of any absolute magnitude and 2) where the detected
fragments must have $H<23$ for consistency with the null result found
in our earlier work \citep{Schunova2012}.

The detectability lifetime decreases as expected as the size of the
progenitor decreases --- smaller progenitors produce even smaller
fragments that are difficult to detect.  The measured value decreases
smoothly from about 11,500 years at $H_p=13$ (about 10\,km diameter)
to about 5,000 years at $H_p=18$ (about 1\,km diameter).  The
detectable lifetime then drops precipitously for families produced by
the smallest progenitors, those with $H_p=20$ that are just barely
capable of producing a detectable family, such that families produced
by progenitors with $H_p=19.5$ can only be identified with our
technique for $\sim 2,000$ years.  This is simply because the smallest
progenitors do not produce enough large fragments to remain above the
survey systems' limiting magnitude.

The probability of actually identifying tidally disrupted NEO families
depends on the relative rates of their production and their lifetime
($L$ or $\tau$) such that the number of detectable families in the
steady state population at any time is $N=FL$ where $F$ is the
production 'flux' rate.  Since the observed number is $N=0$
\citep{Schunova2012} a limit can be placed on the flux rate as
described in \S\ref{sec:TIDI-limits}.

\subsection{End states of tidal disruption fragments}
\label{sec:end-states}

\begin{figure}
\centering
\includegraphics[width=\textwidth]{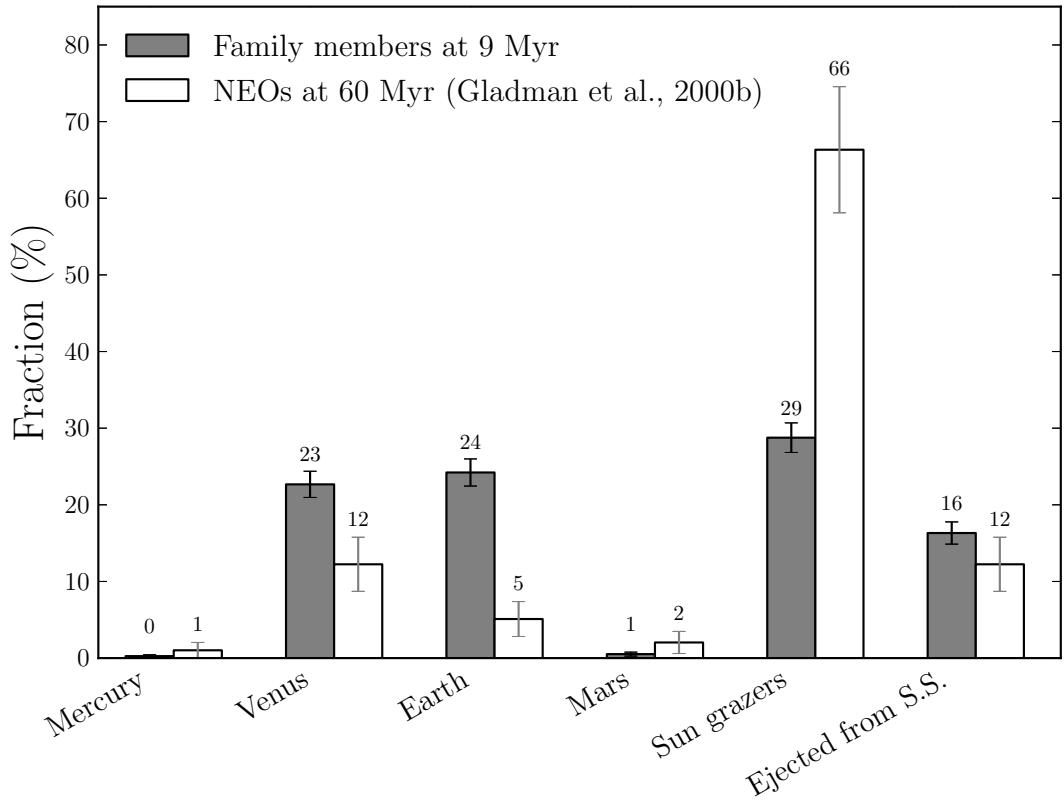} 
\caption{(shaded) End-states of fragments from tidal disruption of
  NEOs at Earth after 9~Myr compared to (white) the dynamical NEO
  population from \citet{Gladman2000} after 60~Myr.  The average
  fraction is provided above the error bar for each bin.}
\label{fig:M-E-collisions-comp}
\end{figure}

About 9~Myr after the s-class tidal disruption at Earth about
50\% of all the fragments are no longer NEOs because they have struck Earth, Venus or the Sun, or were ejected from the solar system (Fig.~\ref{fig:M-E-collisions-comp}).  Roughly one-quarter of all the fragments strike either Earth or Venus while the Sun attracts a slightly higher percentage (though nearly identical to within our statistics).  The fraction of fragments that strike the Sun is smaller than the fraction of the generic NEO population simply because there is a smaller fraction available left to do so after about half of them have already struck Earth and Venus.

\begin{figure}
\centering
\includegraphics[width=\textwidth]{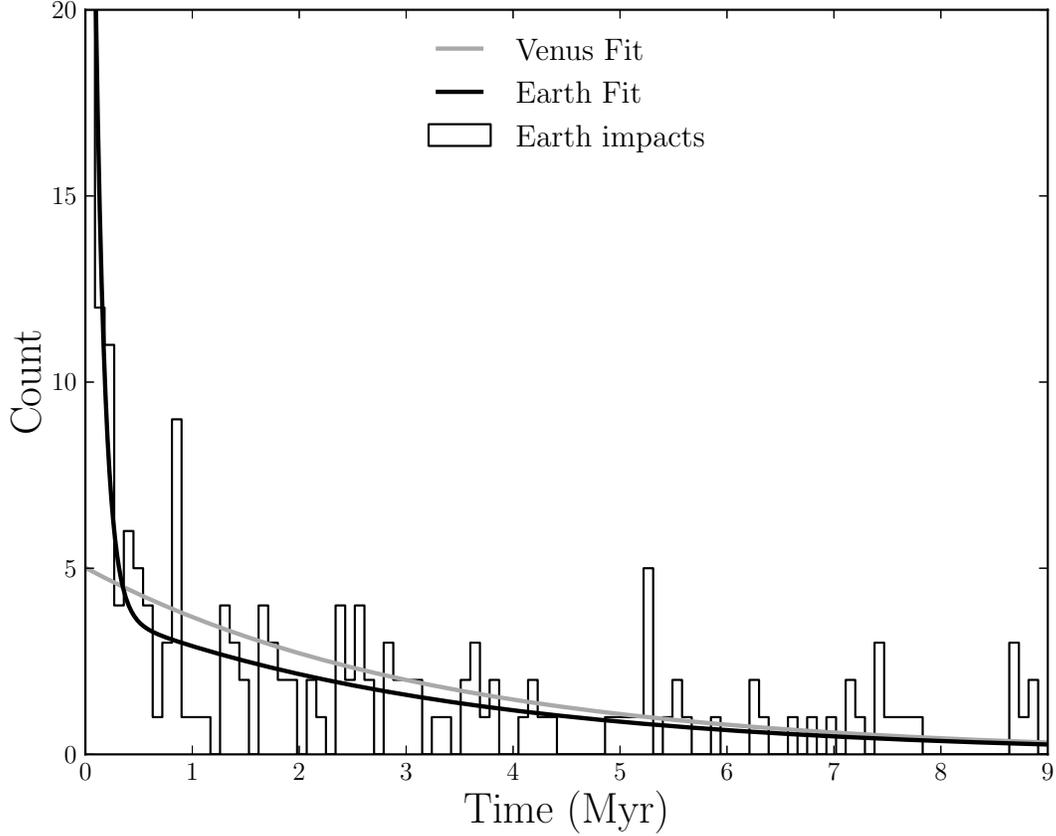} 
\caption{Post tidal distribution collision times of fragments with
  Venus and Earth. The number that collide
  with Earth as a function of time was fit to a double-exponential decay of the form $N_E(t) = N_{fast,E} \exp(t/\tau_{fast,E}) + N_{slow,E} \exp(t/\tau_{slow,E})$ with $\tau_{fast,E} = 0.09\pm0.01$~Myr and $\tau_{slow,E} = 3.3\pm0.7$~Myr.  Note that the first bin with 34 entries is truncated.  The Venus impacts (not shown because they are similar in structure to the Earth impacts) were fit to a single-exponential decay of the form $N_V(t) = N_{slow,V} \exp(t/\tau_{slow,V})$ with $\tau_{slow,V} = 3.3\pm0.4$~Myr.}
\label{fig:E-collisions-distribution}
\end{figure}

It is not surprising that the fraction of fragments that strike Earth is about $5\times$ higher than the fraction of NEOs that strike Earth as measured by \citet{Gladman2000} --- since they were created at Earth they tend to come back to their point of origin.  Indeed, there is a dramatic spike in the impact rate at Earth in the first 100,000~years following a catastrophic disruption (Fig.~\ref{fig:E-collisions-distribution}).  Thus, if an NEO with $D>1\km$ ($H_p\lesssim17.5$) suffers an s-class event near Earth about every 6,000
years \citep{rich1998} then it is likely that Earth is continuously creating shrapnel that increases the impact collision risk above that measured for the dynamical background NEO population.  Every catastrophic disruption produces many $<100$~m
fragments that were not tracked in our work so that each event
 temporarily increases the density of small NEOs on Earth-like orbits for several tens of
thousands of years. Many of them will collide with Earth and cause
localized damage or, more optimistically, they would be good targets for future 
spacecraft missions due to their favorable orbits and low
$v_{\infty}$.

The decay rate for impacts on Venus and Earth are identical after removing the `fast' component due to impacts on Earth in the first 100~kyr.  This `slow' component of the impacts has a decay rate about $30\times$ longer than the fast component with a typical decay time of about 3.3~Myr.  On this time scale the orbits of the fragments have a lot of time to evolve dynamically back into more typical NEO-like orbits.

\section{Discussion}

\subsection{Future improvements}

The coherence times of our simulated families differ by up to an order
of magnitude due to the unique circumstances of each family's
post-disruption orbit.  None of our families remained dynamically
coherent for more than about 70,000 years and, depending on the size
of the progenitor, the detectable lifetime is on the order of
only thousands of years. An extensive quantitative study similar to
this one but using a much larger sample of
progenitors and fully exploring the phase space of rotation rates,
shapes, pole orientations, \etc, is required to draw a better picture of the evolution of tidally created
NEO families.  Such a study might be capable of identifying
regions in NEO orbital element space where a family could remain
clustered for long times and therefore be detectable on longer time
scales.  Indeed, our NEO cluster detection algorithm was applied by
\citet{Schunova2012} over the entire NEO population.  It might be
modifiable to specifically examine the statistical significance of NEO
clusters in the range of orbit elements most likely to contain young
tidally disrupted NEO families.  Combining the orbit element phase space
search for NEO clusters with a taxonomic classification might provide
a more statistically robust detection mechanism
\citep{Ivezic2002,jewitt2006}.

In \S\ref{sec:MOPS} we mentioned that we ignored trailing losses in
our NEO detection simulations and we can justify doing so because the
relatively large $H<23$ NEOs considered in this analysis are usually
detected far enough away that their apparent rate of motion on the sky
plane is small.  Extending our technique to smaller objects in the
several- to hundred-meter size range would require an exquisite
understanding of their detection efficiency as they will only be
detected when they are bright enough during a very close approach to
Earth --- but under these circumstances their rate of motion on the
sky will be high and they can escape detection due to trailing losses
during the exposure time \citep[\eg][]{Veres2012}.  Trailing losses
occur when fast moving asteroids leave trails on the detector so that
the amount of flux per pixel decreases relative to the flux for a
stationary object of the same intrinsic brightness.  For example,
trailing losses start to occur in the \PSone\ survey when objects move
faster than about $0.5$~deg/day \citep{denneau2013}.  Detecting
the small, trailed asteroids in the images will require more
sophisticated image analysis techniques and a good measurement of
their detection efficiency will be required to incorporate the small
objects into an NEO cluster search algorithm.

\subsection{Limits on s-class event frequency}
\label{sec:TIDI-limits}

\begin{figure}
\centering
\includegraphics[width=0.9\textwidth]{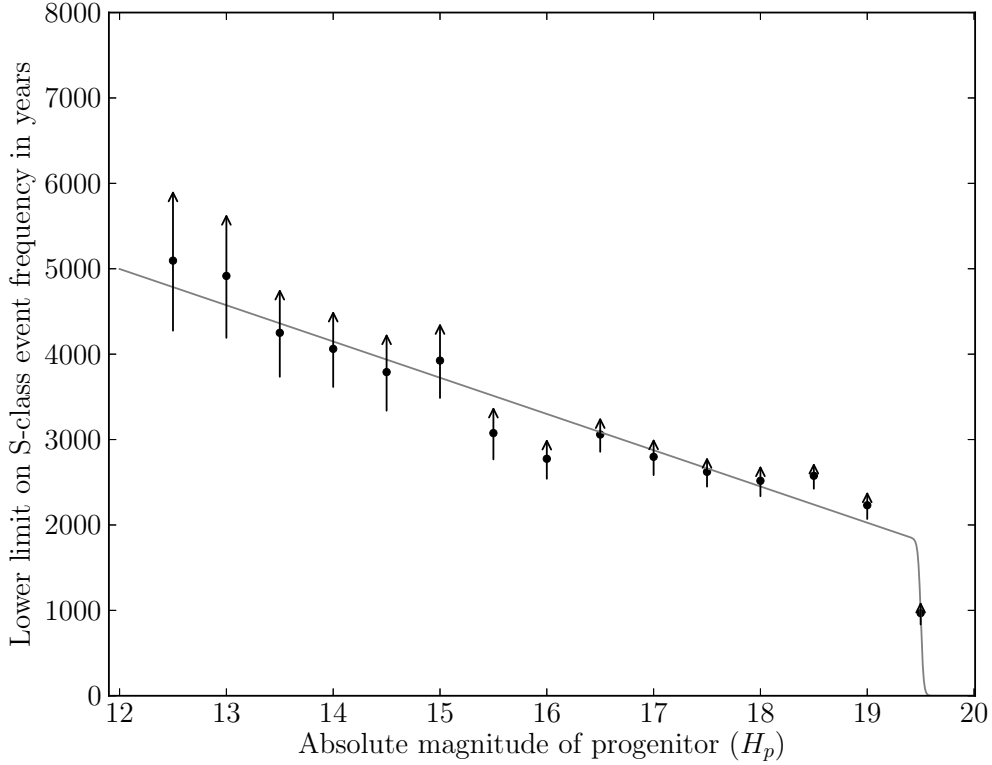} 
\caption{Lower limit on the interval between detectable
  family-producing s-class disruption events as a function of the
  progenitor's absolute magnitude $H_p$.  The central value on each
  data point is the formally calculated limit and the range of the
  arrow indicates the uncertainty on the limit due to the uncertainty
  on the lifetime of families produced at each progenitor's absolute
  magnitude.  The limit was fit to a function of the form $F_S
  = (F_0 + m \, H_p ) / [1 + \exp\{{ (H_p - H_{limit})/ H_{width}}\}]$ with
  $F_0=(10.1\pm0.6) \times 10^3\yr$, $m=-424\pm39\yr$/mag,
  $H_{limit}=19.5$ and $H_{width}=0.1$~mag (we do not provide the
  uncertainties for the last two parameters because they are much
  larger than the precision required to reproduce the fit shown on the
  figure).}
\label{fig:TIDI-frequency-vs-Hp}
\end{figure}

\citet{Schunova2012} detected zero NEO families with fragments of $H<23$
which allows us to set observational limits on the s-class
disruption frequency assuming 1) that all NEO families are created by
s-class tidal disruptions and 2) ignoring the 33\% contribution from
Venus (\ie\ assuming all tidal disruptions occur at Earth;  see \S\ref{sec:Mars-Venus}).  Since the steady state number $N$ of detectable
s-class families in the NEO population is the product of their flux $F$
(creation rate) and average detectable lifetime $L=\bar\tau_{det}$, we
can set an observational lower limit on the creation rate of the
tidally disrupted families $F_S$ (see
Fig.~\ref{fig:TIDI-frequency-vs-Hp}).  

At about the 90\% confidence
level a 1$\km$ NEO s-class disruption can take place no more
frequently than once per $\sim2,300\yr$ or an NEO family would already
have been detected.  We can not place a robust percentage on the
confidence limit because we have not performed enough tidal disruption
simulations with a realistic distribution of physical parameters.
Furthermore, we have so far ignored the detailed contribution of
s-class families to the NEO population by tidal disruption at Venus.
For example, while we have estimated that Venus produces s-class families at
about 33\% of the rate produced by Earth the {\it detectable}
lifetimes of those families is probably considerably shorter due to
their larger mean distance from Earth so they probably make only a
small contribution to the detectable number of NEO families.

Our lower limit on the disruption rate of 1 every $1,000\yr$ at
$H_p=20$ corresponding to objects about $300\meter$ in diameter, is
about 6$\times$ times smaller than the predicted rate for $200\meter$
NEOs of one every 6,200 years as calculated by \citet{toth2011} based
on the SFD of \citet{ivanov2008}.  Our limit is about 3$\times$
smaller than the rate of one $300\meter$ NEOs every 3,100 yr that we
calculate based on the collision probability of NEOs with Earth
\citep{bot1994} assuming that the Earth's impact cross-section is
equivalent to a sphere with a radius 1.5$\times$ the Earth's to allow
for gravitational focussing and that tidal disruptions do not require
impact.  

In other words, it is unlikely that a tidally disrupted NEO
cluster could be detected given the predicted rate and given our
current cluster detection algorithms and observed set of NEOs.  To
detect such a cluster would require either luck, \eg\ a recent tidal
disruption of an NEO, or more sensitive NEO surveys to identify more
and smaller objects.  The problem with detecting more NEOs is that it
increases the number of background objects at the same time as it
increases the number of detected fragments from a disruption event and it is unclear how the \citet{Schunova2012} detection algorithm will
perform in the presence of a
larger background population.

\subsection{Other evidence for disruptions and implications for asteroid space missions}
\label{sec:Space-missions}

It is possible that there already exists evidence of a population of
asteroids created in a tidal disruption in the form of an enhancement
in the number of small objects in Earth-like orbits
\citep{Rabinowitz1993} or the discrepancy between the observed
\citep{mainzer2011} and predicted number \citep{bot2002a} of NEOs
on low-inclination orbits.  Both observations might be explained as
enhancements due to a recent tidal disruption of NEOs by Earth.

\begin{figure}
\centering
\includegraphics[width=0.9\textwidth]{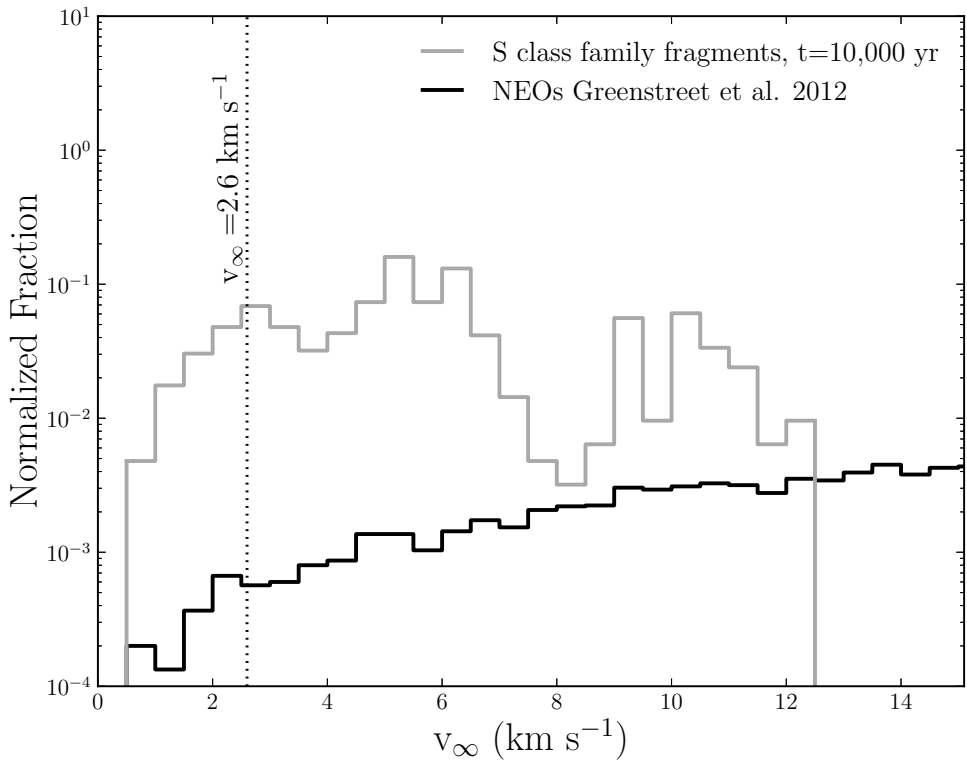} 
\caption{$v_{\infty}$ of all s-class family members 10,000 years after
  tidal disruption compared to the distribution for NEOs from
  \citet{greenstreet2012}. Roughly 10\% of tidal disruption fragments
  have $v_{\infty}<2.6\km\second^{-1}$ as required for NASA's Asteroid
  Retrieval mission compared to just 0.1\% for the background
  (dynamical) NEO population.  The latter distribution has a maximum
  at $v_{\infty}\sim60\km\second^{-1}$.}
\label{fig:V_inf-distribution}
\end{figure}

We will concentrate on the small NEOs with $27<H<31$ in Earth-like
orbits because they are of interest in NASA's Asteroid Retrieval
Mission (ARM) concept that seeks to bring a small heliocentric object
into Earth orbit.  The ARM initiative\footnote{P. Chodas (JPL) -
  personal communication} would like to target objects with $
v_{\infty}<2.6\km\second^{-1}$.  Assuming that we can extend the
average SFD slope of $\alpha=0.55$ for s-class families to the size
range of interest and using the average $\Delta H_1=1$ there are $\sim
2.6 \times 10^{7}$ fragments in the ARM size range produced in the
disruption of a single 1~km parent
body.  Roughly 10\% of the fragments from
an s-class tidal disruption meet the ARM $v_{\infty}$ threshold
compared to only 0.1\% of the overall NEO population (see Fig.~\ref{fig:V_inf-distribution}).  The
$v_{\infty}$ distribution increases with time from disruption as the
members dissipate into NEO orbit element space but many fragments will
still fulfill the mission criterion.  Thus, the tidal disruption of a
single 1~km progenitor at Earth could generate $\sim 3 \times 10^{6}$
ARM targets.  The steady state population is about 1,200 ARM targets
assuming that one 1~km diameter NEO tidally disrupts at Earth every 2,500~years.

Within the dynamical NEO population we expect $\sim4\times10^{8-9}$
NEOs in the same size range depending on the underlying NEO model
\citep[\eg][]{brown2002,Harris2012} of which $\sim0.1$\% satisfy the
ARM $v_{\infty}$ requirement yielding $\sim 4 \times 10^{5-6}$
objects.  The steady-state number of potential ARM targets created by
tidal disruption is too low to explain the observed excess of small
NEOs on Earth-like orbits but a disruption event within the past $\sim
10,000$~years might double the population of ARM targets in the NEO
population and contribute to the explanation of the discrepancy
between the \citet{bot2002a} NEO model and observations such as those
by \citet{Rabinowitz1993} and \citet{mainzer2011}. 

\section{Conclusion}

We have shown that catastrophic tidal disruptions of NEOs as they pass
close to Earth are capable of creating detectable NEO families.  Our simulations suggest that the members of NEO families created by
tidal disruptions are highly porous with a size-frequency distribution
$\propto 10^{(0.55\pm0.11)\,H}$, where the uncertainty is dominated by
the systematic errors in our analysis.  

The rapid dynamical evolution of the members of the tidally disrupted families
results in their rapid dissipation so that they can only be
identified by their orbital similarity for several tens of thousands
of years with an average size-independent family coherence time of
$\tau_{c}=(14.7\pm0.6)\times10^3\yr$.  This value sets an upper limit
on the family's detectable lifetime $\tau_{det}$ --- the time during
which the family members may be detectable by current surveys and
identified as statistically significant clusters in orbit element
space.  The detectability lifetime decreases from about 11,500 years
for progenitors with absolute magnitudes of $H_p=13$ (about 10\,km
diameter) to about 5,000 years at $H_p=18$ (about 1\,km
diameter). This timescale is extremely short compared to main belt
families that are detectable on Gyr timescales and probably explains
why zero NEO families have been discovered to-date.

The fragments from a tidal disruption at Earth have a high probability of impacting Earth at future apparitions within about 100~kyr.  It is possible that the creation rate of this kind of family is high enough that Earth is continuously creating shrapnel that increases the risk of collisions with Earth above the rate expected from the background dynamical NEO population.

The smallest parent body capable of producing an NEO family detectable
with current surveys and our cluster search technique is
$H_{p,max}\sim20$ (for $\bar\alpha\sim0.55$ and $\overbar{\Delta
  H_1}\sim1.0$) corresponding to asteroids of about 0.3 to 0.7~km in
diameter depending on their albedo.

The null detection of NEO families by \citet{Schunova2012} allowed us
to set a lower limit on the frequency of s-class NEO family producing
events. We conclude that at about the 90\% confidence level a $1\km$
NEO can disrupt and create a s-class family no more frequently than
once per $\sim 2,300\yr$, or no more than once per $\sim 1,000 \yr$
for the smallest possible detectable progenitor with $H_{p,max}\sim
20$.  \ie\a real NEO family would have been already detected
\citet[\eg\ ][]{Schunova2012} if the rate were more frequent.  These
limits are a few times smaller than the theoretical expectations so we
have not usefully constrained dynamical or disruption models.
Instead, we have shown that the identification of tidally disrupted
NEO families will be difficult unless new techniques are developed, or
existing techniques are modified to search specifically for tidal
disruptions.  The combination of spectral similarity with orbital
similarity may be fruitful at establishing statistically robust
families under the assumption that the dust produced during the tidal
disruption reaccumulates homegeneously on the fragments.

Nonetheless, every tidal disruption event is capable of producing a
NEO `stream' --- up to several million fragments for a 1~km
progenitor. We speculate that a relatively recent disruption event
that is as-yet-unidentified with an NEO family may have contributed to
the observed excess of objects on low-inclination orbits compared to
the current model predictions.  Such an event would create local
orbit-element density enhancements of small NEOs in Earth-like
orbits that would be excellent targets for future asteroid spacecraft
exploration like the Asteroid Retrieval Mission (ARM) due to their low
$v_{\infty}$ and associated low $\Delta v$.

\clearpage

\section*{Acknowledgments}

This work was supported in part by NASA NEOO grant NNXO8AR22G.  ES's
work was also funded by The National Scholarship Programme of the Slovak
Republic for the Support of Mobility of Students, PhD Students,
University Teachers and Researchers and VEGA grant No. 1/0636/09 from
the Ministry of Education Of Slovak Republic.  MG was funded by
grants \#136132 and \#137853 from the Academy of Finland, and KJW
acknowledges support from NLSI CLOE.


\clearpage

\bibliographystyle{agsm}
\bibliography{tidi}


\clearpage
\begin{figure}
  \centering
  \includegraphics[width=0.9\textwidth]{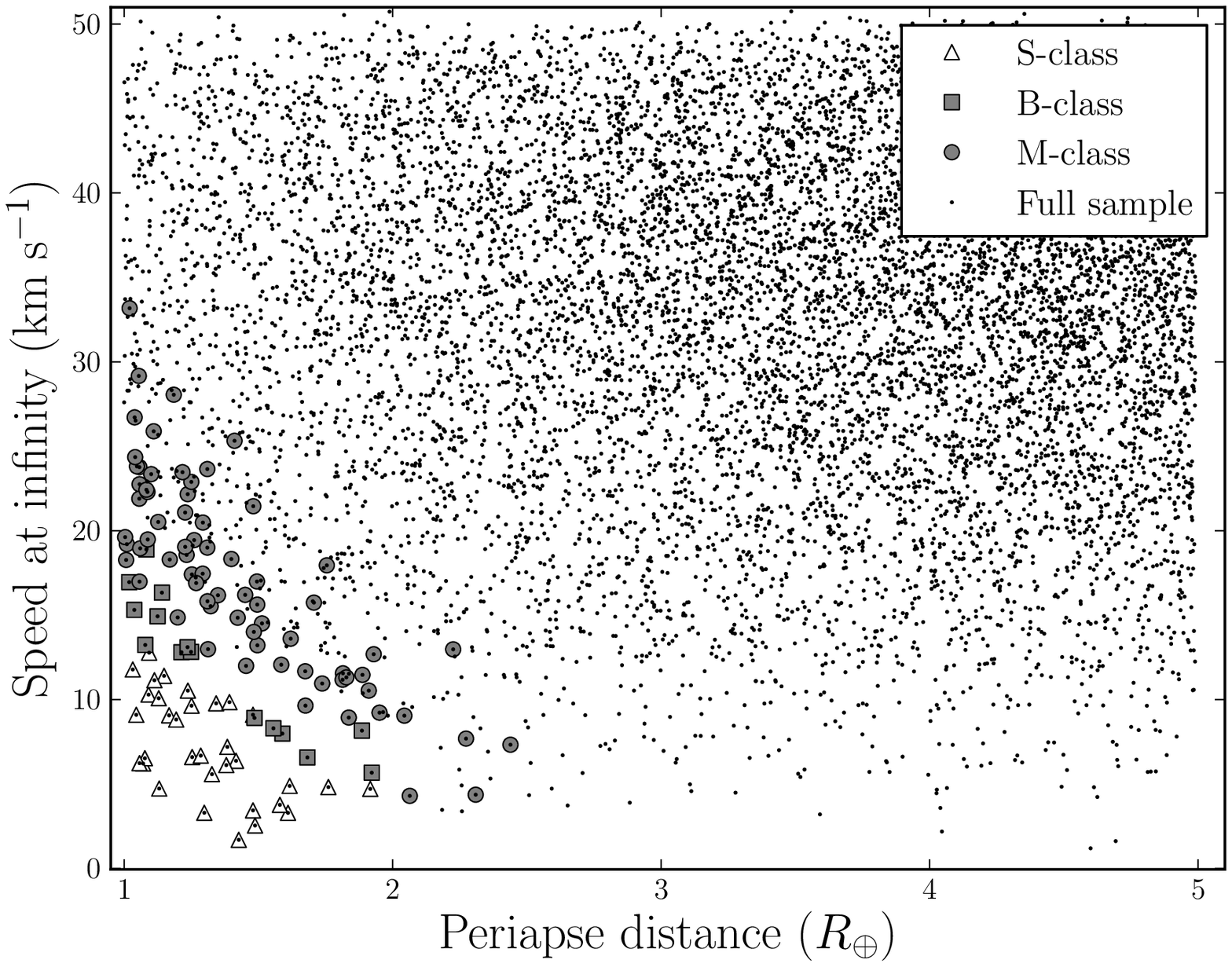}
  \caption{Speed at infinity ($v_{\infty}$) vs. minimum geocentric
    distance (periapse, $q$) for 10,000 synthetic tidal disruption
    progenitors.}
\label{fig:Vinf-q-success-trials}
\end{figure}

\clearpage

\clearpage

\clearpage

\clearpage

\clearpage

\clearpage

\clearpage

\clearpage

\clearpage

\clearpage

\clearpage

\clearpage

\clearpage

\end{document}